\documentclass[
 aps,
 amsmath,amssymb,reprint,
pra,superscriptaddress,floatfix]{revtex4-2}

\usepackage{mathrsfs}
\usepackage{mathtools}
\usepackage{graphicx}%
\usepackage{dcolumn}%
\usepackage{bm}%
\usepackage{pifont} %
\usepackage{siunitx}
\usepackage[normalem]{ulem}
\usepackage[sectionbib]{bibunits}
\usepackage{booktabs}

\DeclareSIUnit\gauss{G}
\DeclareSIUnit{\au}{a.u.}

\usepackage[usenames,dvipsnames]{xcolor}
\usepackage[colorlinks=true,urlcolor=BrickRed,citecolor=BrickRed,linkcolor=BrickRed]{hyperref}
\usepackage[final]{microtype}
\usepackage[dvipsnames]{xcolor}

\usepackage{orcidlink}

\makeatletter
\newcommand{\ORCID}[1]{{\orcidlink{#1}}}
\makeatother

\newcommand{\ket}[1]{\ensuremath{\left|#1\right>}}
\newcommand{\bra}[1]{\ensuremath{\left<#1\right|}}
\newcommand{\ReducedMat}[3]{\ensuremath{\left<\left. #1 \right\| #2 \left\| #3 \right.\right>}}

\newcommand{\etal}{{\it et al.\/}}

\newcommand{\eref}[1]{Equation~(\ref{#1})}
\newcommand{\Eref}[1]{Equation (\ref{#1})}

\newcommand{\fref}[1]{Fig.~\ref{#1}}
\newcommand{\figref}[1]{Fig.~\ref{#1}}
\newcommand{\Fref}[1]{Figure~\ref{#1}}

\newcommand{\lb}[1]{\textbf{#1}}
\newcommand{\flb}[1]{\textbf{#1.}}

\newcommand{\typ}[2]{\texorpdfstring{$J^p={\frac{#1}{2}}^#2$}{J\^p=(#1/2)\^#2}}
\renewcommand{\figurename}{\textbf{Fig.}}
\makeatletter
\def\fnum@figure{\figurename\nobreakspace\textbf{\thefigure}}
\def\@caption@fignum@sep{ {\boldmath $|$} }
\makeatother
\setcitestyle{super}

\makeatletter
\let\oldonlinecite\onlinecite
\renewcommand*{\onlinecite}[1]{%
  \begingroup
    \romannumeral-`\x %
    \NAT@superfalse
    \let\@cite\NAT@citenum
    \oldonlinecite{#1}%
  \endgroup
}
\makeatother

\begin{document}
\makeatletter
\defaultbibliography{\pre@bibdata,References}
\makeatother
\defaultbibliographystyle{trybib}

\preprint{AIP/123-QED}

\newcommand{\paperTitle}{Angular-momentum quantization as intrinsic reference for complete RF polarimetry with Rydberg atoms}
\newcommand{\paperAuthors}{%
\author{Matthew Chilcott\ORCID{0000-0002-1664-6477}}
\affiliation{
Department of Physics, QSO—Quantum Science Otago, and Dodd-Walls Centre for Photonic and Quantum Technologies,
University of Otago, Dunedin 9016, New Zealand
}%
\author{Laurits N. Stokholm\ORCID{0009-0000-5887-7831}}
\affiliation{
Department of Physics and Astronomy,
Aarhus University, Denmark
}%
\author{Matthew Cloutman\ORCID{0009-0009-5746-8425}}
\affiliation{
Department of Physics, QSO—Quantum Science Otago, and Dodd-Walls Centre for Photonic and Quantum Technologies,
University of Otago, Dunedin 9016, New Zealand
}%
\author{J. Susanne Otto\ORCID{0000-0003-0760-3800}}
\affiliation{
Department of Physics, QSO—Quantum Science Otago, and Dodd-Walls Centre for Photonic and Quantum Technologies,
University of Otago, Dunedin 9016, New Zealand
}%
\author{Amita B. Deb\ORCID{0000-0002-2427-3500}}%
\affiliation{
School of Physics and Astronomy, University of Birmingham, Edgbaston, Birmingham B15 2TT, United Kingdom
}%
\author{Niels Kj{\ae}rgaard\ORCID{0000-0002-7830-9468}}%
 \email{niels.kjaergaard@otago.ac.nz}
\affiliation{
Department of Physics, QSO—Quantum Science Otago, and Dodd-Walls Centre for Photonic and Quantum Technologies,
University of Otago, Dunedin 9016, New Zealand
}%
}

\begin{bibunit}
\begin{abstract}
Rydberg electrometry leverages non-metallic vapor cells with highly excited atoms to link RF field properties to a spectroscopic readout in the optical domain. Demonstrations have so far mainly addressed linearly polarized fields inducing a rotationally invariant atomic dressed-level splitting. Here we report on measurements of RF fields in a general state of polarization (SOP) which we map onto the Poincar\'e sphere through spectroscopic fingerprints characteristic of a member of an angular-momentum hierarchy. For SOPs circumnavigating the Poincar\'e sphere's prime meridian, we witness a continuous transformation of the member's eigenenergy spectrum. 
Conversely, such fingerprints can be used as input to a protocol for polarimetric characterization of the RF field, and we demonstrate its extension to SOPs beyond the prime meridian. Because the relative positions of eigenenergies are locked in place by angular-momentum quantization, the framework is universal and calibration-free, and our specific demonstration in rubidium generalizes to all systems with a single valence electron.

\end{abstract}
\title{\paperTitle}
\paperAuthors
\date{\today}
\maketitle
State of polarization (SOP) measurements of electromagnetic fields are used in many scientific, industrial, and technological applications, including material characterization, chemical analysis, and remote sensing. At optical frequencies, a light field’s SOP can be mapped onto the Poincaré sphere using birefringent optical retarders, polarizers, and photodetectors to measure polarization-dependent intensities \cite{Born1999,Collett2005,Hauge1980}. In the RF domain, where, for example, polarization plays an important role in analyzing radar echoes \cite{Drabowitch1998}, polarimetry is typically carried out by means of dual-polarized metallic antennas that measure the amplitude and phase for two orthogonal field components~\cite{Inoue2025}. Due to their conducting nature, such antennas inevitably distort an incoming RF field.
\begin{figure}[bt!]
    \centering
\includegraphics[width=\columnwidth]{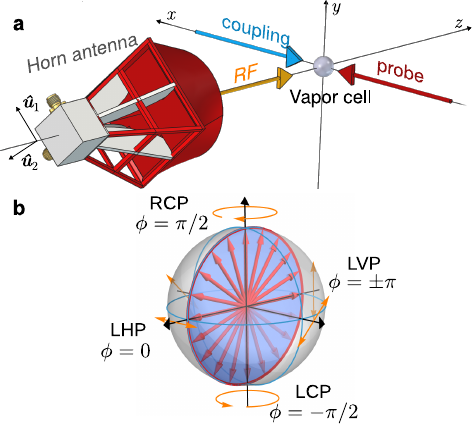}
    \caption{\flb{Rydberg atomic SOP analyzer} \lb{a} Schematic for SOP measurement. The RF field produced by a dual-polarized ($\hat{\bf{u}_1}$/$\hat{\bf{u}_2}$) lensed horn antenna propagates in the $z$-direction towards an atomic vapor cell located at the origin. Coupling and probe laser beams counter-propagate along the $y$-axis. \lb{b} Poincar\'{e} sphere showing the trajectory of the RF-field's Stokes vector (red arrow) when $\phi=-\pi\rightarrow \pi$ in \eref{eq:phi}. As $\phi$ varies, the polarization state evolves from linear-vertical (LVP) to left-circular (LCP) to linear-horizontal (LHP) to right-circular (RCP), and back to LVP.}
    \label{fig:setup}
\end{figure}

Rydberg atomic vapor cells open up the possibility of non-metallic sensors for electromagnetic fields with frequencies falling below the optical domain \cite{Sedlacek2012}. We shall invariably refer to such fields as RF fields even when their frequencies extend into the GHz and THz region. The prospect of SI-traceable and calibration-free RF electric field measurement based on the quantized energy-level structure of atoms~\cite{Holloway2014} has attracted considerable interest from metrology laboratories and beyond~\cite{Schlossberger2024}. Studies concerned with Rydberg atomic sensing of polarization have however mainly been limited to linearly-polarized fields~\cite{Sedlacek2013, Song2018, Wang:23, Cloutman2024,  Yin2024,You2024, Cloutman2026, Cai2025}.

A first demonstration involving elliptically-polarized fields only emerged very recently~\cite{Elgee2024,Elgee2025} and required the operation of no less than three auxiliary local oscillator fields produced by separate microwave antennas in addition to two lasers establishing the electromagnetically-induced transparency (EIT) optical readout scheme, while polarization control has been achieved in cold, spin-polarized ensembles using a bias magnetic field and three phase-controlled microwave sources~\cite{Kurdak2026}.

In this work, we demonstrate that a complete Rydberg atomic RF polarimeter can be implemented without relying on any auxiliary RF fields. Instead, a spectroscopic vapour cell, ``wired up'' with two laser beams  \textit{inherently} processes the device-under-test (DUT) field and maps its SOP onto an optically recorded spectrum.
More specifically, the SOP extracted by the optical interrogation is directly encoded in the quantum states of two Rydberg levels $r_1$ and $r_2$ of an alkali atom with total electronic angular momenta $J$ and $J'$, with projection quantum numbers $m_J$ and $m'_{J'}$. The $J\leftrightarrow J'$-transition can be placed in an iterative hierarchy, where the algebra of quantized angular momentum universally and analytically defines families of SOP-dependent spectroscopic ``fingerprints''.

\begin{figure}[t!]
    \centering
\includegraphics[width=0.95\columnwidth]{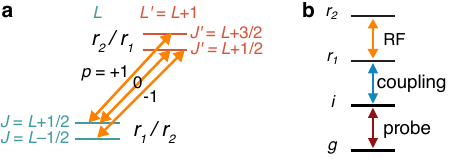}
    \caption{\flb{Atomic level diagrams} \lb{a}  The three dipole-allowed transitions ($L \leftrightarrow L+1$) between the two Rydberg levels $r_1$ and $r_2$ are designated by $p=0,\pm 1$. It is possible for $r_1$ and $r_2$ to assume the role of either the $L$-level or $(L+1)$-level \lb{b} Schematic energy level scheme for ladder-type  EIT-sensing of an RF field showing how the optical probe and coupling  fields connect the atomic ground level $g$ to the RF-field-dressed Rydberg levels $r_1$ and $r_2$ via an intermediate level $i$.}
    \label{fig:level}
\end{figure}
We realize the first four members of the angular momentum transition hierarchy in experiments and elucidate how their associated spectrograms are tied to the SOP and direction of an incoming RF field dressing the atomic states. Hinging on our analytical expressions for the spectroscopic fingerprints, we develop a procedure for determining the SOP of a test field from an experimentally acquired EIT spectrum. This includes judiciously selecting the properties of the optical fields which interrogate the RF-dressed atoms to break the symmetry of potentially duplicate SOP fingerprints to obtain an unambiguous mapping onto the Poincar\'{e} sphere. Since our method is rooted in the quantization of angular momentum and uses this as a metrological ``ruler'', it is independent of the power of the DUT field and it carries no reference to the species of atom involved, including the radial transition matrix elements between its Rydberg states. In this sense the scheme is universal and calibration-free.
\section*{Results}
\subsection*{Transition hierarchy and spectroscopic fingerprints}
\begin{figure*}[h!]
   \centering
\includegraphics[width=\linewidth]{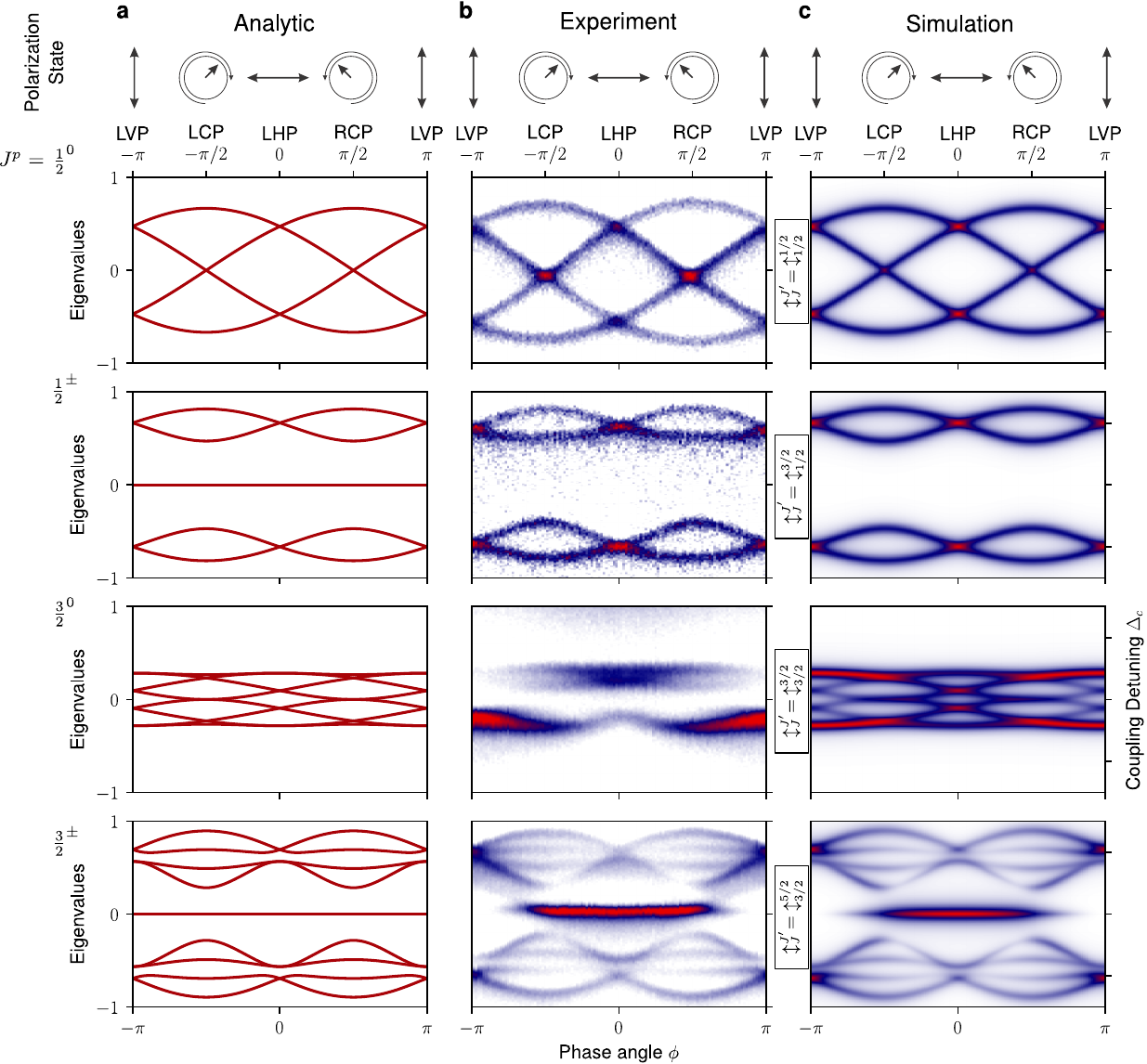}
    \caption{\flb{Spectrograms for RF-dressed Rydberg atomic transitions} Eigenvalue spectra for resonantly coupled manifolds $\updownarrow^{J'}_J=\left\{\updownarrow^{1/2}_{1/2},\updownarrow^{3/2}_{1/2},\updownarrow^{3/2}_{3/2},\updownarrow^{5/2}_{3/2}\right\}$ as a function of the phase angle $\phi$ defined in \eref{eq:phi}. \lb{a} Analytic eigenvalues found by diagonalizing the coupling matrices ${\bf M}^{(Jp)}$ with entries defined by \eref{eq:big}. The SOP of the RF dressing field is shown at the top. \lb{b} Experimentally acquired spectrograms. \lb{c} Simulated spectrograms from density matrix calculations on optically interrogated two-level Rydberg manifolds.}
    \label{fig:zoo}
\end{figure*}
In our demonstration of an optically-wired atomic SOP analyzer, a microwave field that resonantly couples states between $r_1$ and $r_2$ is emitted by a dual-polarized horn antenna [see \fref{fig:setup}a]. The antenna has two input ports, connected to orthogonally oriented stubs in the waveguide region near the horn throat, which we drive with two equal-amplitude phase-synchronous sinusoidal signals of frequency $f$. Each stub generates a linearly-polarized field of amplitude $E_0/\sqrt{2}$ propagating on the axis of the horn, which we take to be along $\boldsymbol{\hat{z}}$. Denoting the two orthogonal directions of polarization as $\boldsymbol{\hat{u}}_1$ and $\boldsymbol{\hat{u}}_2$, respectively, the resultant electric component of the emitted field can be expressed as 
\begin{equation}
    \boldsymbol{E}(z,t)=\frac{E_0}{\sqrt{2}}\operatorname{Re} \left[(\boldsymbol{\hat{u}}_1+e^{i\phi}\boldsymbol{\hat{u}}_2)e^{i(kz-\omega t)}\right], \label{eq:phi}
\end{equation}
where $k=2\pi f/c$, and $\phi$ describes the phase of the second signal source with respect to the first. By adjusting the phase $\phi$ we can control the polarization state of the microwave field impinging on the atoms to reside anywhere on the prime meridian of the Poincar\'{e} sphere shown in \fref{fig:setup}b, with ellipticity and helicity given by
\begin{equation}
    \epsilon=\frac{|\sin\phi|}{1+|\cos
    \phi|},
\end{equation}
and
\begin{equation}
    h={\rm sgn}\left(\tan\phi/2\right),
\end{equation}
respectively.
By rotating the RF antenna about the $z$-axis we can access any Poincar\'{e} sphere meridian. Conversely, for an SOP of any RF polarization azimuth (ellipse orientation angle) it is possible to rotate the coordinate system in \fref{fig:setup}a so this becomes zero and the Stokes vector lies in the prime meridian plane (blue plane in \fref{fig:setup}b). Because of this, we shall first concern ourselves with SOPs on the prime meridian before turning to the problem of determining the appropriate rotation from spectroscopic measurements.

In our experiment, the atomic sensing cell is located at the origin of a Cartesian coordinate system, with the antenna producing the test field oriented such that $\boldsymbol{\hat{u}}_{1,2}=(\boldsymbol{\hat{x}}\pm\boldsymbol{\hat{y}})/\sqrt{2}$. For $\phi=0$, the resultant field is linearly polarized along $\boldsymbol{\hat{x}}$---the horizontal direction of our setup 
[cf. \fref{fig:setup}a]. Expressed in terms of covariant spherical-basis unit vectors $\boldsymbol{\hat{e}}_{\pm 1}=\mp(\boldsymbol{\hat{x}}\pm i\boldsymbol{\hat{y}})/\sqrt{2}$, an atom located at the origin will experience an RF field
\begin{align}\label{eq:Efield}
\boldsymbol{E}(t) = \frac{E_0}{\sqrt{2}}\, \operatorname{Re} \bigg\{ 
\bigg[ & -\left( \cos\frac{\phi}{2} - \sin\frac{\phi}{2} \right) \boldsymbol{\hat{e}}_{+1} \notag \\
      +\; & \left( \cos\frac{\phi}{2} + \sin\frac{\phi}{2} \right) \boldsymbol{\hat{e}}_{-1} 
\bigg] e^{-i(\omega t-\phi/2)} \bigg\}.
\end{align}
By choosing $\boldsymbol{\hat{z}}=\boldsymbol{\hat{e}}_0$ as the quantization axis for the atomic system, and working in the dipole and rotating wave approximations, the angular part of the coupling Hamiltonian, which describes allowed transitions between $r_1$ and $r_2$, can be represented analytically by a matrix $\bm M^{(Jp)}$ of dimension $(2J+2J'+2)\times(2J+2J'+2)$ with entries 

\begin{equation}
\begin{aligned}
M_{ij}^{Jp}
={}&
\sqrt{\frac{2J+1}{4(J+1)}}
\sqrt{\frac{(2J+3)^{|p|}}{J^{1-|p|}}}
\\
&\times\sum_{q=\pm1}q\Biggl[
\begin{aligned}[t]
&(-1)^j\,
\text{\small$\begin{pmatrix}
J & 1 & J+|p|\\
J+1-j & q & i-3J-|p|-2
\end{pmatrix}$}
\\
&\mathllap{{}+{}}(-1)^i\,
\text{\small$\begin{pmatrix}
J & 1 & J+|p|\\
J+1-i & q & j-3J-|p|-2
\end{pmatrix}$}
\Biggr]
\end{aligned}
\\
&\qquad\times
\left(\cos\frac{\phi}{2}-q\sin\frac{\phi}{2}\right),
\end{aligned}
\label{eq:big}
\end{equation}
where $p=0,\pm1$ is defined in \fref{fig:level}a.
In deriving \eref{eq:big}, we have made use of the condition $J'-J=0,1$; for coupling between the levels to be dipole-allowed $\Delta J=0,\pm1$ is required \cite{Sobelman1992} and we take $J'\geq J$ without loss of generality. We have also assumed an alkali atom with electronic spin quantum number $S=1/2$.\nocite{Sobelman1992}
For example, the coupling matrix for the case $J^p=\frac{1}{2}^0$ takes the appearance

\begin{equation}
\boldsymbol{M}^{\frac12\,0}
=
-\frac{\sqrt{2}}{3}
\left(
\vcenter{\hbox{%
\scriptsize
\setlength{\arraycolsep}{2pt}%
$\begin{array}{cccc}
0 & 0 & 0 &
\cos\!\!\frac{\phi}{2}\!{+}\!\sin\!\!\frac{\phi}{2}
\\
0 & 0 &
\cos\!\!\frac{\phi}{2}\!{-}\!\sin\!\!\frac{\phi}{2}
& 0
\\
0 &
\cos\!\!\frac{\phi}{2}\!{-}\!\sin\!\!\frac{\phi}{2}
& 0 & 0
\\
\cos\!\!\frac{\phi}{2}\!{+}\!\sin\!\!\frac{\phi}{2}
& 0 & 0 & 0
\end{array}$%
}}
\right),
\label{eq:repr}
\end{equation}
with the ordering of basis states in play, \ket{J,m_J} and \ket{J',{m}'_{J'}} given in Methods.

By diagonalising $\bm M^{(Jp)}$ it is possible to find analytic expressions for the variation in eigenvalues with $\phi$. The simplest member of the iterative hierarchy established by \eref{eq:big}, \typ{1}{0} [with the explicit interaction matrix  \eref{eq:repr}],  gives the four angular-dependent eigenvalues 
\begin{equation}
\lambda_n=\frac{2}{3}\operatorname{Re}\left(\exp\Big\{i\big[\phi/2+(2n-1)\pi/4\big]\Big\}\right),\ n=1\dots 4\label{eq:eigenvalues}
\end{equation}

Moving up in the hierarchy increases the complexity with the number of eigenvalues growing linearly with $J$. \Fref{fig:zoo}a presents the angular variation in eigenenergies for the cases of the coupled manifolds $\updownarrow^{J'}_J=\left\{\updownarrow^{1/2}_{1/2},\updownarrow^{3/2}_{1/2},\updownarrow^{3/2}_{3/2},\updownarrow^{5/2}_{3/2}\right\}$, corresponding to $J^p=\left\{\frac{1}{2}^0,\frac{1}{2}^\pm,\frac{3}{2}^0,\frac{3}{2}^\pm\right\}$( Supplementary Note~\ref{sec:sup:atlas} shows the $J\rightarrow\infty$ limiting behaviour). In accordance with the rule for linearly polarized RF fields given in Ref.~\nocite{Cloutman2024}\onlinecite{Cloutman2024}, for $\phi=0$ (LHP) and $\phi=\pi$ (LVP), the number of unique eigenvalues is $N_{\rm eig}^{\rm lin}=2,3,4,5$, respectively, with all eigenvalues doubly degenerate.  For elliptical polarization ($\phi\neq 0,\pi$), the degeneracies of the non-zero eigenvalues are lifted by the chirality acquired by the field, when its SOP moves off the equator of the Poincar\'{e} sphere. If a  zero eigenvalue occurs at linear polarization (only and always for $J^{\pm}$ members), it remains doubly degenerate for all $\phi$ for its two eigenstates, which lie entirely in the $J'$ manifold. The maximum number of unique eigenvalues that can be encountered in our cases is therefore $N_{\rm eig}^{\rm max}=4,5,8,9$, respectively. 

\subsection*{Experiments}
The above treatment leading to \fref{fig:zoo}a elucidates how the SOP of the DUT field is imprinted on the energy spectrum of the dressed states hybridized from substates of $r_1$ and $r_2$. To demonstrate how these analytically predicted eigenenergy fingerprints can be used to infer the polarization state of the RF dressing field, we conduct experiments on $\rm ^{87}Rb$ atoms in a vapor cell. Specifically, we make use of a $\sim18$~GHz DUT RF field and consider its effect on Rydberg states with principal quantum numbers around 50, but the results established will be of general validity as the scheme is based on universal results on quantized angular momentum structure.

The RF-dressed manifold of atomic energies is probed optically via an electromagnetically-induced transparency (EIT) ladder scheme (see \fref{fig:level}b).  Details of our detection scheme have been described elsewhere~\cite{Cloutman2024, Cloutman2026}. In brief, a beam of $\sim780$~nm probe laser light along the $y$-axis addresses a transition from the Rb ground level $5S_{1/2}$ to an intermediate $5P_{3/2}$-level, which is further connected to a Rydberg level $r_1$ by a $\sim480$~nm counter-propagating coupling laser (see Fig.~\ref{fig:setup}a) with some detuning $\Delta_c$. In the default operation of our optically-interrogated sensor, both probe and coupling light fields are linearly polarized in the $\boldsymbol{\hat{x}}$-direction. When $\Delta_c$ is scanned over a $\sim200$~MHz range, EIT peaks will emerge in the detected probe field whenever the coupling laser resonantly links a state of the intermediate level ($i$) to an $r_1$ state or the $r_1$ component of a dressed state hybridized from the two Rydberg levels. 

In Fig.~\ref{fig:zoo}b we show spectrograms resulting from EIT-probing the dressed manifolds $\left\{\updownarrow^{59P_{1/2}}_{59S_{1/2}},\updownarrow^{59P_{3/2}}_{59S_{1/2}},\updownarrow^{49D_{3/2}}_{50P_{3/2}},\updownarrow^{49D_{5/2}}_{50P_{3/2}}\right\}$ as $\phi$ is scanned from $-\pi$ to $\pi$. In doing so, the Stokes vector describing the field-SOP circumnavigates a meridian on the Poincar\'{e} sphere (cf.~Fig.~\ref{fig:setup}b). The four Rydberg transitions are particular incarnations of $J^p=\left\{\frac{1}{2}^0,\frac{1}{2}^+,\frac{3}{2}^0,\frac{3}{2}^+\right\}$. Figure~\ref{fig:zoo}c shows the result of density matrix simulations \cite{Cloutman2026,Chilcott2026a}, which unlike the analytic results incorporate effects pertaining to the optical fields (Methods). We note how the simulated spectra replicate the peak positions of the analytical predictions in Fig.~\ref{fig:zoo}a, while also capturing the change in prominence of the moving spectral features as $\phi$ is varied. It is important to note that the SOP of the RF field entirely defines the relative positions of spectral peaks. However, for a given value of $\phi$, the observed intensity for each peak of the spectrum will depend on properties of the optical fields such as polarization and propagation direction.
\begin{figure}[t!]
\includegraphics[width=0.75\linewidth]{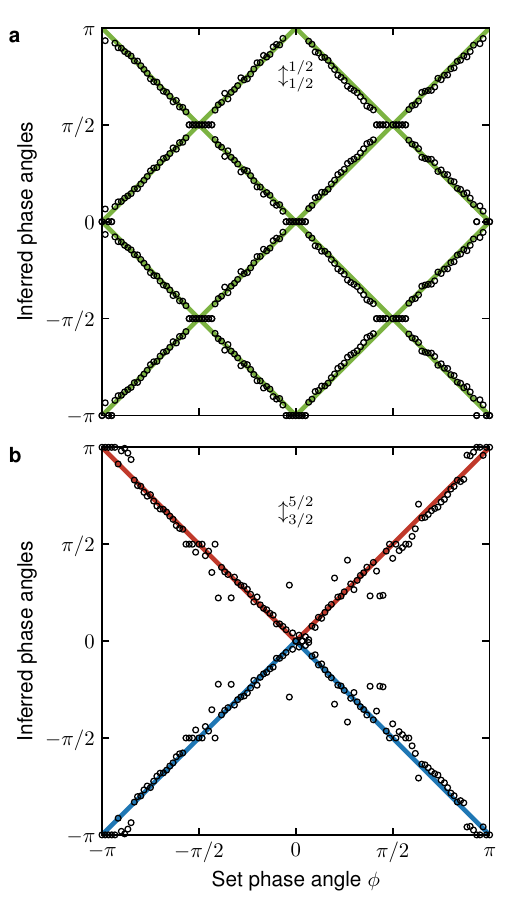}
\caption{\flb{RF field SOP inferred from EIT spectra} Possible phase angles extracted from experimentally acquired EIT spectra of \fref{fig:zoo}b as a function of a set phase angle $\phi$. The extraction relies on determining the eigenvalue envelopes for a given spectrum (Methods and Supplementary Note~\ref{sec:sup:envelope}
). {\bf a} $\updownarrow^{1/2}_{1/2}$-Rydberg transition showing a fourfold ambiguity in the inferred phase angle. {\bf b} $\updownarrow^{5/2}_{3/2}$-Rydberg transition for which the ambiguity in the inferred phase angle is twofold as $\phi$ and $-\phi$ give rise to the same EIT spectral fingerprint (see \fref{fig:zoo}b).\label{fig:inversion}}
\end{figure}
\begin{figure}
    \includegraphics[width=0.7\linewidth]{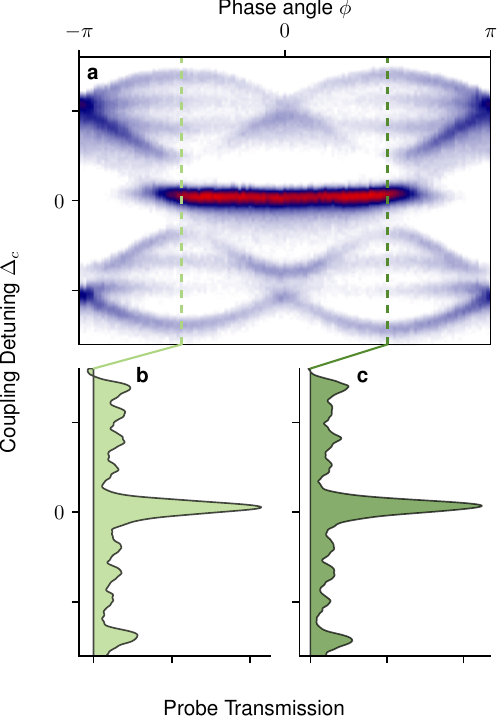}
    \caption{\flb{EIT spectra for atoms dressed with circularly polarized RF fields} \textbf{a} Experimental spectrogram for a field-dressed $\updownarrow^{5/2}_{3/2}$-transition probed in the configuration \fref{fig:setup}a with dashed light and dark green lines marking the sections of LCP ($\phi=-\pi/2$) and RCP ($\phi=\pi/2$) RF-fields. \textbf{b,c} The nominally identical spectra recorded for LCP and RCP RF-fields. \label{fig:90}}
\end{figure}

In discussing our experimental results in Fig.~\ref{fig:zoo}b, we consider first the simplest case of a $\updownarrow^{59P_{1/2}}_{59S_{1/2}}$-transition (i.e. \typ{1}{0}) for which we find excellent agreement between experiment and the analytically predicted pattern of Fig.~\ref{fig:zoo}a which replicates Ref.~\onlinecite{Chopinaud2024}. In particular we note how pure linear and circular RF polarizations give rise to exactly two and three spectral peaks, respectively, with the maximum four peaks attained for elliptical polarization. Meanwhile, the simulated spectrum captures the increase in peak intensity when eigenvalue degeneracy occurs. From the four eigenvalues that can be extracted from a given measured \typ{1}{0} spectrum we can extract the corresponding phase angle analytically (Methods). \Fref{fig:inversion}a shows the result of carrying out the inversion from spectrum to phase angle. The mirror symmetries about integer multiples of $\pi/2$ for the \typ{1}{0}-spectrogram
introduce a four-fold ambiguity when inferring $\phi$ from a given spectrum. For example, $\phi=\pm\pi/4,\pm 3\pi/4$ display identical spectra. We stress that, although the individual peak positions depend on the RF power, all splittings scale linearly with the applied electric field amplitude. Because our inversion uses only ratios of splittings (see Methods), the common scaling factor cancels, and the result is independent of the power of the DUT field.

\begin{figure*}[t!]
\includegraphics[width=0.9\linewidth]{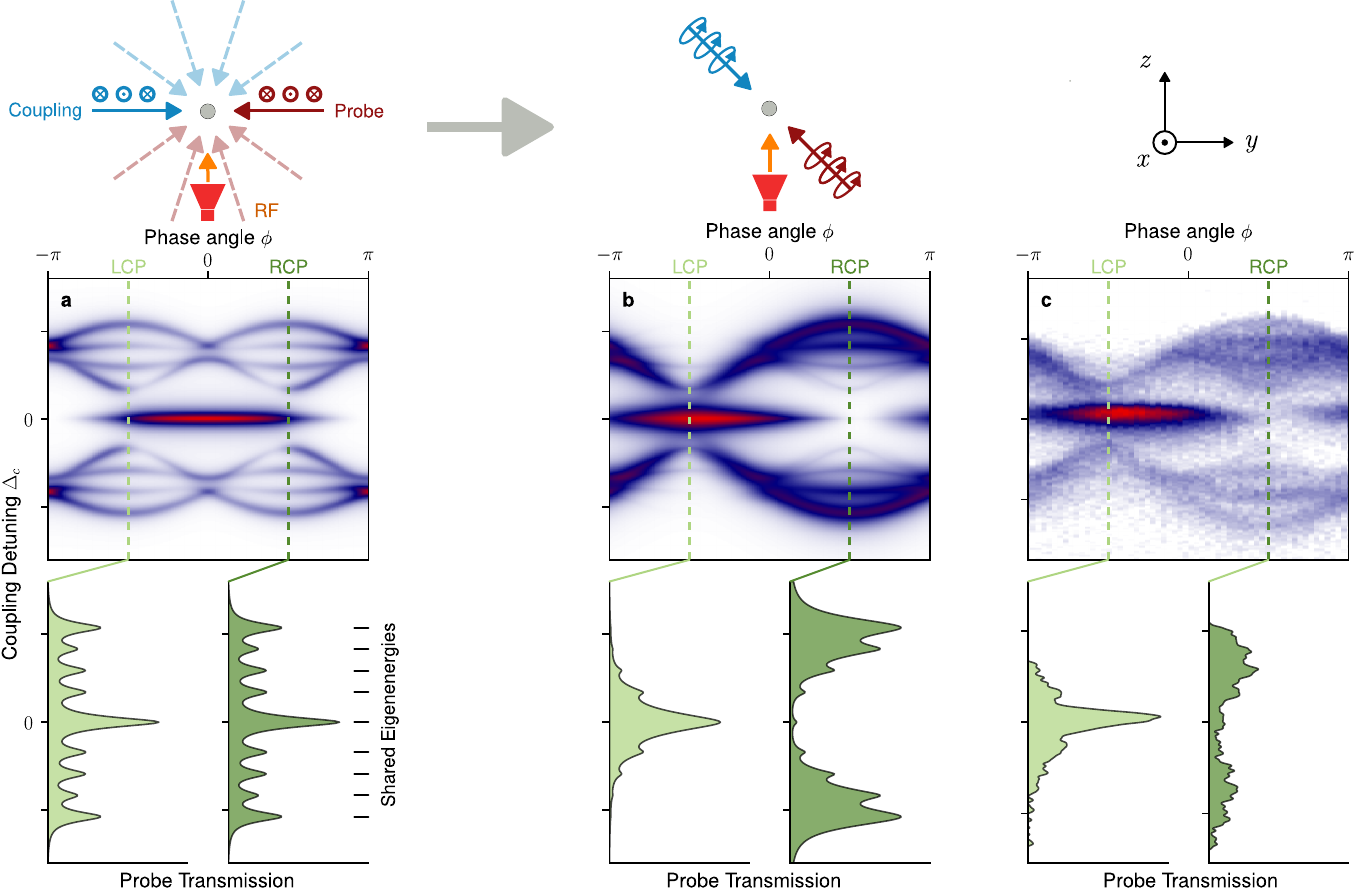}
\caption{\flb{Breaking spectrogram symmetry} Effect of altering the optical configuration when EIT-probing a field-dressed $\updownarrow^{5/2}_{3/2}$-transition. \lb{a} Simulated spectrogram for optical fields     linearly polarized along $\bf \hat{x}$ and counter-propagating in the $yz$-plane; the spectrogram is invariant under a rotation of the optical beam direction about $\bf \hat{x}$. The spectra shown in the lower panels are indistinguishable. \lb{b} Simulated spectrogram for circularly polarized optical probe and coupling beams propagating along $\bf \hat{y}-\hat{z}$. From the optical EIT spectra shown in the lower panel it is clearly possible to discriminate between LCP and RCP RF dressing fields. In particular, the central peak is more visible when the RF field has the same handedness as the probe field. \lb{c} Experimental realization of the optical configuration of {\bf b}. }\label{fig:45}
\end{figure*}
For the dressed $59S_{1/2}\leftrightarrow 59P_{3/2}$ transition (i.e., \typ{1}{+}) the agreement with the analytical prediction is also good. For the experimental spectrum, we note an absence of a central peak at $\Delta_c=0$ so that the number of coupling matrix eigenenergies exceeds the number of spectral peaks by one everywhere: as noted above, zero-eigenvalue states lie entirely in the $J'$ manifold, which in this case ($59P_{3/2}$ level) means they have exclusively $r_2$-character and the coupling field therefore does not (Laporte's rule) connect to them from the $i$ level (see \fref{fig:level}b). The absence of a central peak is captured by the density matrix simulations along with an increase in peak intensity for degenerate eigenenergies.

\begin{figure*}[tb!]
\includegraphics[width=0.9\linewidth]{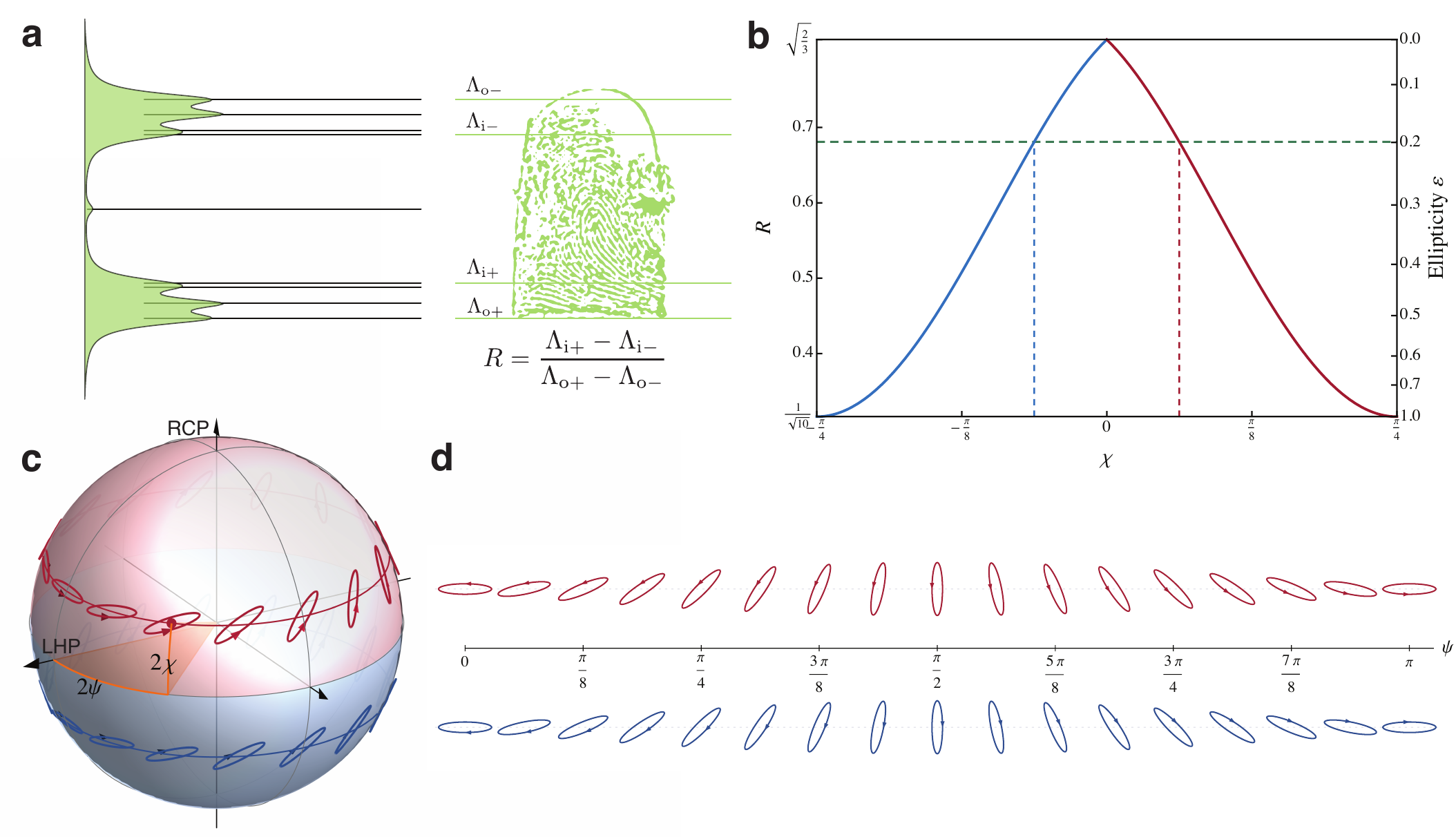}
\caption{\flb{Spectroscopic fingerprints} \lb{a} EIT spectrum for a \typ{3}{+} system displaying 9 dressed state eigenvalues. The outer and nonzero inner eigenvalues define the envelope ratio $R$ (see Methods), which uniquely determines the ellipticity $\epsilon$ of the RF DUT dressing field. \lb{b} Eigenvalue-envelope ratio $R$ as a function of Poincar\'{e} sphere ellipticity angle $\chi$ (see \lb{c} for definition). A given value of $R$ corresponds to two equal-magnitude ellipticity angles of opposite sign. \lb{c} Poincar\'{e} sphere showing the definition of the ellipticity angle $\chi$ and the polarization azimuth $\psi$. Polarization states on the red and blue parallels shown have the same ellipticity but opposite helicity (captured by the sign of $\chi$). Along each parallel, the polarization ellipse rotates (\lb{d}) such that its major axis makes an angle $\psi$ with respect to horizontal.}\label{fig:newfig}
\end{figure*}
The spectral SOP ambiguity is reduced when proceeding to the cases of $J=3/2$ where experiment and density-matrix simulation reveal a simple mirror symmetry about $\phi=0$.  The recorded experimental spectrum for the dressed $\updownarrow^{49D_{3/2}}_{50P_{3/2}}$ transition, however, neither resembles the analytical \typ{3}{0} prediction in Fig.~\ref{fig:zoo}a nor the density matrix simulation in Fig.~\ref{fig:zoo}c. The discrepancy results from dealing with a real atom and is caused by interference from the nearby $49D_{5/2}$ level compromising the idealized case of an isolated $\updownarrow^{49D_{3/2}}_{50P_{3/2}}$-transition. Supplementary Note~\ref{sec:sup:3level} details the effect of adding a third level to the problem while also showing that agreement between experiment and simulation is established by expanding the density matrix to include the $49D_{5/2}$ state.

In contrast, the $\updownarrow^{49D_{5/2}}_{50P_{3/2}}$-transition is well described within a simple two-level framework and excellent correspondence between analytical theory, experiment, and simulations is found. For a set phase angle $\phi$ of the test field, \fref{fig:inversion}b shows the phase angles that result from inverting $\updownarrow^{49D_{5/2}}_{50P_{3/2}}$-spectra. Similarly to the $\updownarrow^{{1/2}}_{{1/2}}$-case, the procedure (Methods) involves calculating a ratio between the inner and outer eigenvalue envelope for a spectrum, and in the first instance it produces four potential solutions, but by taking into account the $\phi$-dependent prominence of the central $\Delta_c=0$-peak, two of these solutions can be discarded. For the purpose of demonstrating Rydberg atomic SOP measurements, we are left with the problem of ascertaining which of the two remaining solutions represents the phase angle of the DUT field. We shall therefore focus our attention on the case of \typ{3}{+} and discuss strategies to establish an unambiguous assignment to the upper or lower Poincar\'{e} hemisphere. 
\subsection*{Resolving helicity ambiguity for a \typ{3}{+} system}
The mirror symmetry about $\phi=0$ for the \typ{3}{+}spectrogram of Fig.~\ref{fig:zoo}b implies a twofold spectral degeneracy for the probing configuration Fig.~\ref{fig:setup}a. Indeed, Fig.~\ref{fig:90} shows the nominally identical  \typ{3}{+} spectra for circularly polarized RF fields of opposite helicities (i.e., $\phi=\pm\pi/2$) on the basis of which it is not possible to discern the handedness of the DUT field.

 To obtain an RF-field SOP measurement, which can discriminate between the two branches  of \fref{fig:inversion} (red and blue lines) the mirror symmetry about $\phi=0$ encountered in the spectrogram of Fig.~\ref{fig:90} needs to be broken. This can be achieved by altering the configuration of the laser beams interrogating the atoms.
 Figure~\ref{fig:45}a,b shows the effect of rotating the probe and coupling beam directions so that they counter-propagate along $\bf \hat{y} - \hat{z}$, while changing their polarizations to be circular. This operation will not change the peak positions in the spectrograms as these are locked in place by the SOP of the RF dressing field. It will, however, modify the prominence of peaks. Specifically, we see how the central peaks in the spectra for $\phi=-\pi/2$ and $\phi=\pi/2$ become enhanced and suppressed, respectively. \Fref{fig:45}c shows that this is indeed what we find in experiment. Probing in the configuration of \fref{fig:45}b therefore reveals the helicity of a circularly polarized RF field and, more generally, it can remove the ambiguity between the two branches of \fref{fig:inversion}b.

 \subsection*{Complete RF polarimetry}

Having demonstrated how to infer the SOP for a Stokes vector with its tip on the prime meridian running through LHP and LVP on the Poincar\'{e} sphere (see \fref{fig:setup}b), we turn to the problem of ascertaining the SOP located on a general meridian of the Poincar\'{e} sphere and we will describe a step-by-step protocol for how this can be done. 

 \Fref{fig:newfig} outlines the general scenario. From a spectroscopic fingerprint acquired in the probe configuration of \fref{fig:setup}a  we can obtain the eigenvalue envelope ratio  $R$ (\fref{fig:newfig}a), which uniquely determines the ellipticity $\epsilon$ (\fref{fig:newfig}b). For a polarization state with a given $\epsilon$ there are, however, two possible helicities (blue and red branches in \fref{fig:newfig}b). Moreover, for each helicity-branch, the state's polarization ellipse can be rotated by any angle from 0 to $180^\circ$ (\fref{fig:newfig}c). The angle $\psi$  between the ellipse's major axis  and the horizontal is called the polarization azimuth (\fref{fig:newfig}d). Hence states with $\psi=0$ and $\psi=\pi/2$ both lie on the prime meridian.
 
Starting from the probing configuration in \fref{fig:setup}a, where the DUT field propagates along the $z$-axis in the horizontal plane, our protocol locates the polarization azimuth by rotating the propagation axis of the counter-propagating probe and coupling beams about $\hat{\bf z}$. We shall designate the angle of this optical axis with respect to the horizontal plane $\alpha$, and, consequently, the configuration in \fref{fig:setup}a corresponds to $\alpha=0$.

For each rotation angle $\alpha$ of the optical axis, one can record an EIT spectrum, and we shall refer to a series of such spectra with evolving $\alpha$ as an $\alpha$-scan. In our current experimental setup we are unable to rotate the optical axis of our probe and coupling beams in the laboratory frame
.
To underpin the description of our protocol with experimental data, we shall therefore emulate $\alpha$-scans by rotating the RF antenna about its axis ($\hat{\mathbf{z}}$), which is easy to do by mounting it on a rotation stage. We stress that the physics involved depends only on the relative rotation: starting from the configuration in Fig.~\ref{fig:setup}a, a rotation of the laser beams' optical axis about $\hat{\mathbf{z}}$ and a rotation of the emitting antenna in the opposite direction by the same amount will lead to the same measurement outcome.

\begin{figure}
    \includegraphics[width=\linewidth]{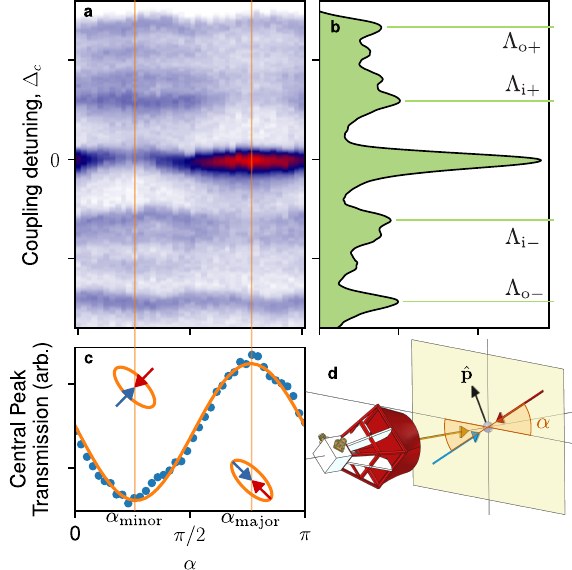}
    \caption{\flb{Extracting $|\chi|$ and $\psi$ from an $\alpha$-scan} \textbf{a} $\alpha$-scan for an example DUT field: a collection of EIT spectra (transmission vs. $\Delta_c$) acquired as $\alpha$ is varied from 0 to $\pi$. \textbf{b} EIT spectrum averaged over $\alpha$. $|\chi|$ can be extracted from the eigenvalue envelope (green lines). \textbf{c} Central peak amplitude as a function of $\alpha$. Location of the  maximum (minimum) occurs when the optical axis of the probe and coupling fields is along the major (minor) axis of the RF field's polarization ellipse. \textbf{d}, Geometry of $\alpha$-scan (cf. \figref{fig:setup}). $\alpha$ is the relative rotation angle of the probe and coupling fields' optical axis about the propagation vector of the RF field.  \label{fig:opticalaxscan}}
\end{figure}

    \Fref{fig:opticalaxscan} shows the result of an $\alpha$-scan for an example DUT field. As the optical axis rotates relative to the incoming DUT field along $\hat{\bf z}$ we see how the positions of eigenvalues remain constant---of course, the RF-dressed eigenenergies are themselves not affected by how we choose to interrogate. The prominence of each eigenvalue peak will however have an $\alpha$-dependence and change with the probing configuration. In particular, the prominence of the central peak provides a key to unlock the polarization azimuth $\psi$ determination. The central peak will namely be maximised when the optical axis is aligned along the major axis of the DUT field's polarization ellipse, while it is minimized when aligned along the minor axis. We designate the two angles for which this happens $\alpha_{\rm major}$ and $\alpha_{\rm minor}$, respectively; of course, $\alpha_{\rm major}=\alpha_{\rm minor}+\pi/2$. For the example data presented in \fref{fig:opticalaxscan}, we can infer that the DUT field has the polarization azimuth $\psi\sim3\pi/4$.

We note that the determination of $\psi$ from an $\alpha$-scan does not rely on the particular shape of the curve traced out in \fref{fig:opticalaxscan}c, but only on its maximum or minimum (or, preferably, both). For a linearly polarized DUT RF field it has been known since Ref.~\onlinecite{Sedlacek2013} that the central EIT peak when using linearly co-polarized optical fields will be at its minimum when all fields are co-polarized and its maximum when the polarizations of the RF and optical fields are perpendicular. The effect of the $\alpha$-rotation at the location of the atoms is to rotate the direction of oscillation for a linearly polarized optical field with respect to RF-polarization. Moreover, a linearly polarized RF field constitutes the limit case of a DUT field with $\epsilon=0$ and we therefore see that maximum EIT transmission corresponds to alignment of the optical axis along the polarization of the DUT field's  `major axis', where the optical and rf polarizations are perpendicular. This result naturally extends to DUT fields with $\epsilon>0$.

Of course, from the eigenvalue envelope ratio $R$ of any of the spectra that make up \Fref{fig:opticalaxscan}a we can infer the ellipticity $\epsilon$ or equivalently $|\chi|$. \Fref{fig:opticalaxscan}b shows the EIT spectrum that results from carrying out an average over $\alpha$, and using this for our determination of $R$ we obtain $|\chi|=\pi/8$ for the DUT field (cf. \fref{fig:newfig}b).

Having determined $\psi$ and $|\chi|$, all that remains for locating the DUT field on the Poincar\'{e} sphere is to determine the sign of $\chi$ which captures the helicity. We have already outlined a procedure for determining the helicity for states on the meridian shown in \fref{fig:setup}b, and shown that this can be used to discriminate between RCP and LCP RF fields (see \fref{fig:45}). Building on this insight, the sign of $\chi$ can be found for a general SOP. 

\begin{figure}
    \includegraphics[width=\linewidth]{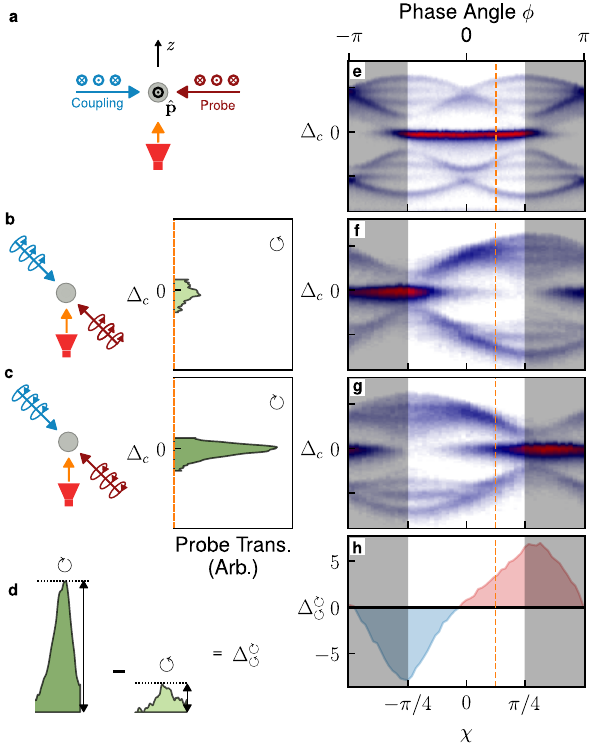}
    \caption{\flb{Protocol for determining the sign of $\chi$}
      \textbf{a} Reference configuration, in which the probe and coupling beams' common optical axis has been rotated about $\hat{\mathbf{z}}$ to $\alpha=\alpha_{\rm major}$, with $\alpha_{\rm major}$ inferred from an $\alpha$-scan of the DUT field (see \fref{fig:opticalaxscan}c,d).
\textbf{b} Configuration ($\circlearrowleft$) obtained from the reference configuration (\lb{a}) by a $45^\circ$ rotation of the optical axis about $\hat{\mathbf{p}}$, with LHC probe and RHC coupling light, and the resulting EIT spectrum around the central peak for the example DUT field of \fref{fig:opticalaxscan}.
\textbf{c} As \textbf{b}, but configuration ($\circlearrowright$) has RHC probe and LHC coupling light.
\textbf{d} The difference in central-peak height between \textbf{c} and \textbf{b} defines $\Delta_{\circlearrowleft}^{\circlearrowright}$, whose sign gives the sign of $\chi$; for the example field $\Delta_{\circlearrowleft}^{\circlearrowright}>0$, hence $\chi>0$.
\textbf{e}--\textbf{g} Spectrograms obtained in configurations \textbf{a}--\textbf{c}, respectively, when scanning $\phi:-\pi\rightarrow\pi$. The dashed orange line marks the example DUT field ($\phi=\pi/4$), at which the spectra in \textbf{b}--\textbf{d} are taken.
\textbf{h}, $\Delta_{\circlearrowleft}^{\circlearrowright}$ as a function of the phase angle $\phi$ (top axis). The corresponding ellipticity angle is $\chi=\frac{1}{2}\sin^{-1}(\sin\phi)$, so that $\chi=\phi/2$ for $|\phi|\leq\pi/2$. Shaded regions, $|\phi|>\pi/2$, correspond to SOPs whose major axis is vertical in the rotated frame. \label{fig:handedness}}
\end{figure}

\Fref{fig:handedness} gives a formalised protocol for determining $\mathrm{sgn}(\chi)$ for an arbitrary SOP on the Poincar\'e sphere. From the result of an $\alpha$-scan, such as the one shown in Fig.~\ref{fig:opticalaxscan}c, we set $\alpha=\alpha_{\rm major}$. This operation is equivalent to a coordinate transformation which shifts the DUT field's SOP along its Poincar\'e-sphere parallel, placing it on the RCP--LHP--LCP half of the prime meridian (setting $\alpha=\alpha_{\rm minor}$ would instead place it on the RCP--LVP--LCP half), while conserving the sign of $\chi$.

By acquiring EIT spectra in the configurations shown in \fref{fig:handedness}b,c, and determining the difference in central-peak height between the two spectra, $\Delta^\circlearrowright_\circlearrowleft$, we obtain $\mathrm{sgn}\,\chi = \mathrm{sgn}\,\Delta^\circlearrowright_\circlearrowleft$ (\fref{fig:handedness}d). \Fref{fig:handedness}h shows the measured variation in $\Delta^\circlearrowright_\circlearrowleft$ when scanning $\phi$ over a $2\pi$ interval. Indeed, we see that $\mathrm{sgn}\,\chi(\phi)$ correlates robustly with $\mathrm{sgn}\,\Delta^\circlearrowright_\circlearrowleft(\phi)$. We note that $\Delta^\circlearrowright_\circlearrowleft$ changes sign in very close proximity of $\chi=0$, where the RF SOP is linear and the field has zero net helicity.

\begin{figure}[t!]
\includegraphics[width=\linewidth]{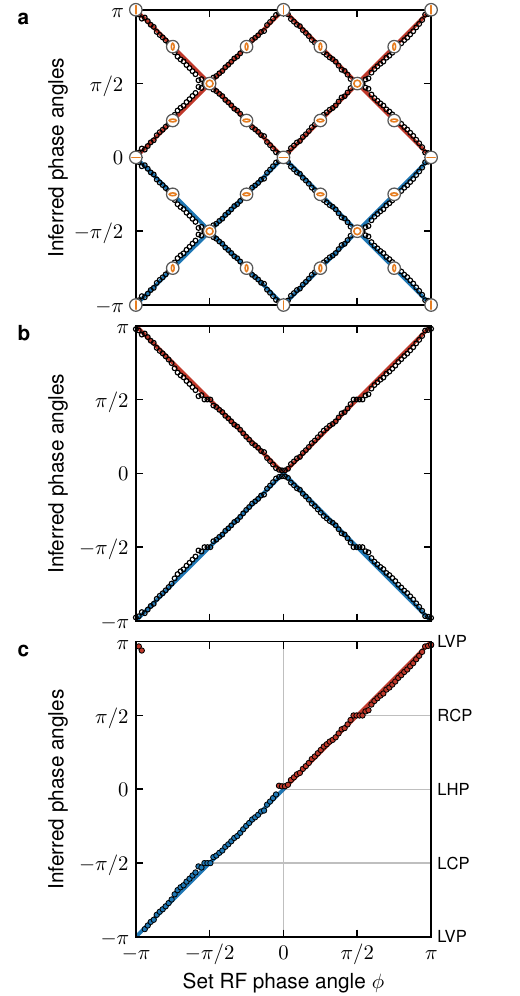}
\caption{\flb{RF field SOP from $\alpha$-averaged EIT spectra} Phase angles inferred by acquiring an $\alpha$-scan for RF DUT fields on the prime meridian of the Poincar\'{e} sphere. \lb{a} Applying the procedure of \fref{fig:newfig}a,b to the $\alpha$-averaged spectra we obtain the ellipticity $\epsilon$ for a set phase $\phi$; the $\epsilon$-fingerprint inverts to up to four phase values. Orange ellipses annotate the orientation (horizontal or vertical) of the major axis on a branch; red (blue) lines denote a positive (negative) value of $\chi$. \lb{b} Selecting branches of \lb{a} according to the spectroscopically inferred $\alpha_{\rm major}$. Compared to \fref{fig:inversion}b, the $\alpha$-scanned data provide a higher fidelity inversion with a $2^\circ$ median deviation; the two-fold helicity ambiguity remains. \lb{c}  Using the protocol of \fref{fig:handedness} a negative (positive) helicity is assigned to each DUT field and shown as blue (red) points. \label{fig:inversion_alpha}}
\end{figure}

Carrying out an $\alpha$-scan is generally also conducive to obtaining more accurate values for $|\chi|$ than what can be obtained from a spectrum at a single value of $\alpha$: because the prominence of a given spectral peak may change with $\alpha$, an $\alpha$-averaged EIT spectrum such as \fref{fig:opticalaxscan}b facilitates a high fidelity extraction of eigenvalues. \Fref{fig:inversion_alpha}a shows the result of inferring $|\chi|$ from $\alpha$-averaged spectra, while \fref{fig:inversion_alpha}b makes use of the inferred $\alpha_{\rm major}$ from the protocol step of \fref{fig:opticalaxscan}c to appropriately select branches with a corresponding ellipse orientation. The fidelity of the inverted phase shows a significant improvement compared to $\alpha=0$ data of \fref{fig:inversion}b. Finally, \fref{fig:inversion_alpha}c uses the protocol step of \fref{fig:handedness} to assign a helicity for each candidate pair and in effect obtain $\operatorname{sgn}{\chi}$. This works well for all states but those very close to linear polarization, where $\Delta^\circlearrowright_\circlearrowleft\approx 0$ and a small amount of measurement noise can affect the assignment.

  \section*{Discussion}
 This work has demonstrated how the SOP for an incoming RF DUT field imprints on the Rydberg states of atoms in a vapor cell sensor and that this imprint can be read out via optical spectroscopy.  In general, a recorded optical spectrum will only reveal the ellipticity $\epsilon=\tan|\chi|$ of the DUT field, but by acquiring multiple spectra at suitable choices of the propagation directions and polarizations of the interrogating optical fields, it is possible to assign a sign to $\chi$, and, moreover, to ascertain the polarization azimuth $\psi$ as illustrated in the case of a dressed $\updownarrow^{5/2}_{3/2}$ transition. In particular, we describe and emulate a detailed protocol for uniquely mapping the polarization state of an RF field onto the Poincar\'{e} sphere.

Our approach is complementary to recent work in cold, spin-polarized ensembles, where a bias magnetic field spectrally resolves the $\sigma^-$, $\pi$ and $\sigma^+$ components, whose amplitudes are set by interfering several phase-controlled microwave sources \cite{Kurdak2026}. There, measurements in spin-unpolarized ensembles were stated to be limited to linearly polarized fields or to require a reference polarization. In our work, which uses an unpolarised room-temperature gas, the angular-momentum structure of the degenerate Rydberg manifolds fixes how the RF polarization components are encoded in the dressed state spectrum. Hence, remarkably, a room-temperature vapour cell without bias field or auxiliary RF fields suffices to characterise an RF field that is not under the experimenter's control---the defining task of a polarimeter.

A hallmark of our protocol is that none of its steps requires calibration against a reference field. The ellipticity $|\chi|$ is retrieved from an eigenvalue ratio that is fixed by angular-momentum algebra, and is independent of the DUT field strength, the radial matrix elements, and the atomic species. The polarization azimuth $\psi$ and the sign of $\chi$ follow from symmetry alone: since the polarization ellipse is mirror-symmetric about its major and minor axes, the extrema of an $\alpha$-scan necessarily coincide with these axes, whatever the detailed shape of the curve; and since reversing the handedness of both optical fields mirrors a reversal of the RF helicity, $\Delta^\circlearrowright_\circlearrowleft$ is odd in $\chi$ and vanishes for linear polarization. The only remaining input---which extremum corresponds to the major axis, and which sign of $\Delta^\circlearrowright_\circlearrowleft$ to which helicity---is fixed once and for all by the dipole selection rules. This contrasts with heterodyne-based Rydberg polarimetry, where the mapping from measured signals to field components is established by calibrating local-oscillator fields against a known reference polarization~\cite{Elgee2024}.

The framework we present is moreover universal: because the spectroscopic fingerprints are determined solely by the angular momenta $J$ and $J'$ of the coupled Rydberg levels, the same hierarchy of fingerprints, and hence the same inversion protocol, applies to any atom or atom-like system with a single valence electron. Supplementary Note~\ref{sec:sup:cs} shows how our inversion scheme can be applied straightforwardly to cesium spectra of the $65D_{5/2} \leftrightarrow 66P_{3/2}$ transition at $\sim 2.5$~GHz, that recently became available~\cite{Schlossberger2026}.The $\alpha$-scan procedure of \fref{fig:opticalaxscan}  is not applicable to $J^p=\frac{1}{2}^0$ spectra, because for any given SOP the prominence of  a spectral peak will be invariant with $\alpha$ \cite{Chopinaud2024,Cloutman2024}, and it is therefore not possible to extract the polarization azimuth. Meanwhile, the ellipticity $\epsilon$ for a $J^p=\frac{1}{2}^0$ spectroscopic fingerprint  can be obtained directly using the inversion \eref{eq:Rinv} as demonstrated in \fref{fig:inversion}a for the rubidium $\updownarrow^{59P_{1/2}}_{59S_{1/2}}$ transition and also in recent work on $\updownarrow^{61P_{1/2}}_{61S_{1/2}}$ using loss spectroscopy in a magneto-optical trap \cite{ElBekai2026}. Carrying out the operation of \fref{fig:45}a,b it is also possible to assign the field's helicity, but the orientation will remain unknown for this lowest member of the angular momentum hierarchy. In contrast, our experiments with $J^p=\frac{3}{2}^+$ demonstrate the feasibility of locating the SOP on the Poincar\'{e} sphere within $\sim2^\circ$ of accuracy.

 As a natural extension of our work, we envisage the directions of the optical beams to be scanned in conjunction with scanning their states of polarization. Such a tomographic approach might be used to compensate for the effect of a possible background magnetic field and other symmetry-breaking perturbations. It may also be used to infer an unknown direction of arrival (DoA) for the incoming RF field---a core diagnostic within RF antenna engineering~\cite{Chen2010}. While conventional DoA estimation typically makes use of an array of antennas~\cite{Molaei2024} to infer a transmitter location, Rydberg atomic sensors offer a solution for carrying out this task based on a single detection element~\cite{Robinson2021,Schlossberger2025,Talashila2025a}. In recent experiments, we have demonstrated that the framework presented in this article allows for a particularly simple implementation of a Rydberg atomic DoA sensor which like our SOP measurements exploits atoms processing an incoming DUT field via their quantized angular momentum structure.
 
\section*{Methods}
\subsection*{Elements of the coupling matrix} The coupling matrix $\boldsymbol{M}$ describes electric-dipole
coupling between magnetic substates of the manifolds $J$ and $J'$.
In the ordered basis
\[
\bigl\{
\ket{J,-J},\ldots,\ket{J,J},
\ket{J',-J'},\ldots,\ket{J',J'}
\bigr\},
\]
its entries are given by \eref{eq:big}, with the column and row
indices specifying the initial and final substates, respectively.
For example, $M^{(\frac{1}{2}0)}_{32}$ describes the coupling
\[
\ket{J=\tfrac12,m=\tfrac12}
\rightarrow
\ket{J'=\tfrac12,m'=-\tfrac12}
\]
for the \typ{1}{0} system. Explicitly, the matrix element is
\[
\bra{J'=\tfrac12,m'=-\tfrac12}
-\boldsymbol{d}\cdot\boldsymbol{E}
\ket{J=\tfrac12,m=\tfrac12},
\]
where $\boldsymbol{d}=-e\boldsymbol{r}$ is the electric-dipole
operator. Expressing $\boldsymbol{E}$ in terms of its spherical
components $E_{\pm1}$ [cf. \eref{eq:Efield}] and applying the Wigner--Eckart theorem,
we obtain
\begin{align}
M^{(\frac12\,0)}_{32}
&\propto \sum_{q=\pm1} E_q
\bra{\tfrac12',-\tfrac12} r_{-q}
\ket{\tfrac12,\tfrac12}
\nonumber\\
&\propto \sum_{q=\pm1} E_q
\begin{pmatrix}
\tfrac12 & 1 & \tfrac12 \\
\tfrac12 & -q & \tfrac12
\end{pmatrix}
\propto E_{+1}
\propto \cos\tfrac{\phi}{2}-\sin\tfrac{\phi}{2},
\end{align}
which up to a multiplicative factor (see Supplementary Note \ref{sec:sup:rq_matrix}) matches the corresponding entry in the matrix \eref{eq:repr} from the generalized expression \eref{eq:big}.

\subsection*{Inference of the phase angle \texorpdfstring{$\phi$}{phi}}

Given an optically acquired spectrum of RF-dressed atoms, we identify the location of the two outermost peaks $\Lambda_{\rm o\pm}$ and two innermost peaks $\Lambda_{\rm i\pm}$ and form the ratio
\begin{equation}\label{eq:ratio}
	R=\frac{\Lambda_{\rm i+}-\Lambda_{\rm i-}}{\Lambda_{\rm  o+}-\Lambda_{\rm o-}}.
\end{equation}
We note that $\pm(\Lambda_{\rm i+}-\Lambda_{\rm i-})/2$ and $\pm(\Lambda_{\rm o+}-\Lambda_{\rm o-})/2$ correspond to, respectively, the two numerically smallest and two numerically largest eigenenergies of the dressed system. These define inner and outer eigenvalue envelopes (Supplementary Note~\ref{sec:sup:envelope}).
\newline
\textbf{Dressed $\updownarrow^{1/2}_{1/2}$ transition.} The analytical expressions for the eigenvalues in the \typ{1}{0} case give $\lambda_{1}=-\lambda_{4}=\frac{2}{3}\sin(\phi/2+\pi/4)$ and $\lambda_{2}=-\lambda_{3}=\frac{2}{3}\cos(\phi/2+\pi/4)$ [see \eref{eq:eigenvalues}]. From \eref{eq:ratio} it follows that $0\leq R\leq 1$. Given an experimentally determined eigenvalue envelope ratio, the phase angle solutions can be obtained by solving the equations $
\tan(\phi/2+\pi/4)= \pm R$ and $\cot(\phi/2+\pi/4)= \pm R$ on the interval $[-\pi,\pi]$. Hence
\begin{equation}
    \phi=\pm\tilde{\phi},\pm(\pi-\tilde{\phi})\label{eq:phis}
\end{equation}
where
\begin{equation}
    \tilde{\phi}=2\left[\frac{\pi}{4}-\tan^{-1} R\right].\label{eq:Rinv}
\end{equation}
\newline
\textbf{Dressed $\updownarrow^{5/2}_{3/2}$ transition.} For the \typ{3}{\pm} case, the exact analytical eigenvalue expressions are somewhat involved (Supplementary Note~\ref{sec:sup:envelope}). However, the envelopes bounding the positive and negative eigenvalues are extremely well approximated as the sum of a simple harmonic function and a constant. The phase angle solutions can be found by solving a quadratic equation in $|\sin\phi|$. The four solutions are given by \eref{eq:phis} with
\begin{equation}
\begin{aligned}
\tilde{\phi}=\arcsin\!\Bigl[
&\sqrt{2-\sqrt{6}\,R+(4-\sqrt{15})R^2}\\
&-\left(\sqrt{5/2}-\sqrt{3/2}\right)R
\Bigr].
\end{aligned}\label{eq:phitilde}
\end{equation}
    Here the eigenvalue envelope ratio is bounded as $\sqrt{1/10}\leq R\leq \sqrt{2/3}$.

\subsection*{Density matrix calculations}
As in Ref.~\citenum{Cloutman2026}, the simulations presented here are extracted from calculations of the steady state of the atomic density matrix in the optical and RF fields. As the probe transition resolves the hyperfine structure, the simulations are performed in the $\ket{F, m_F}$ basis, and include a dummy state to handle (effective) non-radiative decay processes. The Python computer code used for these density matrix simulations is provided in the Supplementary Material of Ref.~\citenum{Cloutman2026}.
\section*{Data Availability}
\noindent The data that support the findings of this study are available from the corresponding
author upon reasonable request.
\section*{Code Availability}\noindent The Python code used for carrying out the density matrix simulations in this work is available in the Supplemental Material of Ref.~\citenum{Cloutman2026}.

\putbib
\end{bibunit}
\section*{ACKNOWLEDGEMENTS}
\noindent N.K. acknowledges feedback from the participants of the symposium celebrating the 25th anniversary of the 5th Institute of Physics, University of Stuttgart, and the 60th birthday of Tilman Pfau (September 2025), where these results were first presented. This work was supported by the Marsden Fund of New Zealand (Contract No. UOO2421) and the Catalyst Fund of the New Zealand Ministry of Business, Innovation \& Employment through Quantum Technology Aotearoa.
\section*{Author Contributions}
\noindent N.K. conceived the project; M.Chilcott and L.N.S. performed experiments with support from M.Cloutman, J.S.O. and A.B.D.; M.Chilcott analysed the data with support from L.N.S. and N.K.; M.Chilcott performed density matrix calculations; M.Chilcott and N.K. made the figures; N.K. wrote the manuscript with support from M.Chilcott and comments from all authors; N.K. supervised the project.
\section*{Competing Interests}The authors declare no competing interests.

\newpage
\pagebreak
\clearpage
\begin{bibunit}
\title{Supplementary Information for: \texorpdfstring{\\}{} \paperTitle}
\paperAuthors

\maketitle
\par\null\clearpage
\onecolumngrid

\setcounter{equation}{0}
\setcounter{figure}{0}
\setcounter{table}{0}
\setcounter{page}{1 }
\renewcommand{\theequation}{S\arabic{equation}}

\makeatletter
\renewcommand{\thesection}{\arabic{section}}

\newif\if@emptysectitle

\let\orig@section\section
\renewcommand{\section}{\@ifstar{\orig@section*}{\@dblarg\my@section}}
\def\my@section[#1]#2{%
  \def\@tempa{#2}%
  \ifx\@tempa\@empty
    \@emptysectitletrue
  \else
    \@emptysectitlefalse
  \fi
  \orig@section[#1]{#2}%
}

\def\@seccntformat#1{%
  \def\@tempa{#1}\def\@tempb{section}%
  \ifx\@tempa\@tempb
    SUPPLEMENTARY NOTE \thesection
    \if@emptysectitle\else:\fi
  \else
    \csname the#1\endcsname.%
  \fi
  \quad}
\makeatother

\renewcommand{\figurename}{\textbf{Supplementary Fig.}}
\renewcommand{\tablename}{\textbf{Supplementary Table}}
\makeatletter
\def\fnum@figure{\figurename\nobreakspace\textbf{\thefigure}}
\def\@caption@fignum@sep{ {\boldmath $|$} }
\renewcommand{\citenumfont}[1]{S#1}   %
\renewcommand{\place@bibnumber}[1]{[S#1]} %
\makeatother

\section{Angular eigenvalue variation}\label{sec:sup:atlas}

\begin{figure}[b!]
    \includegraphics[width=0.8\linewidth]{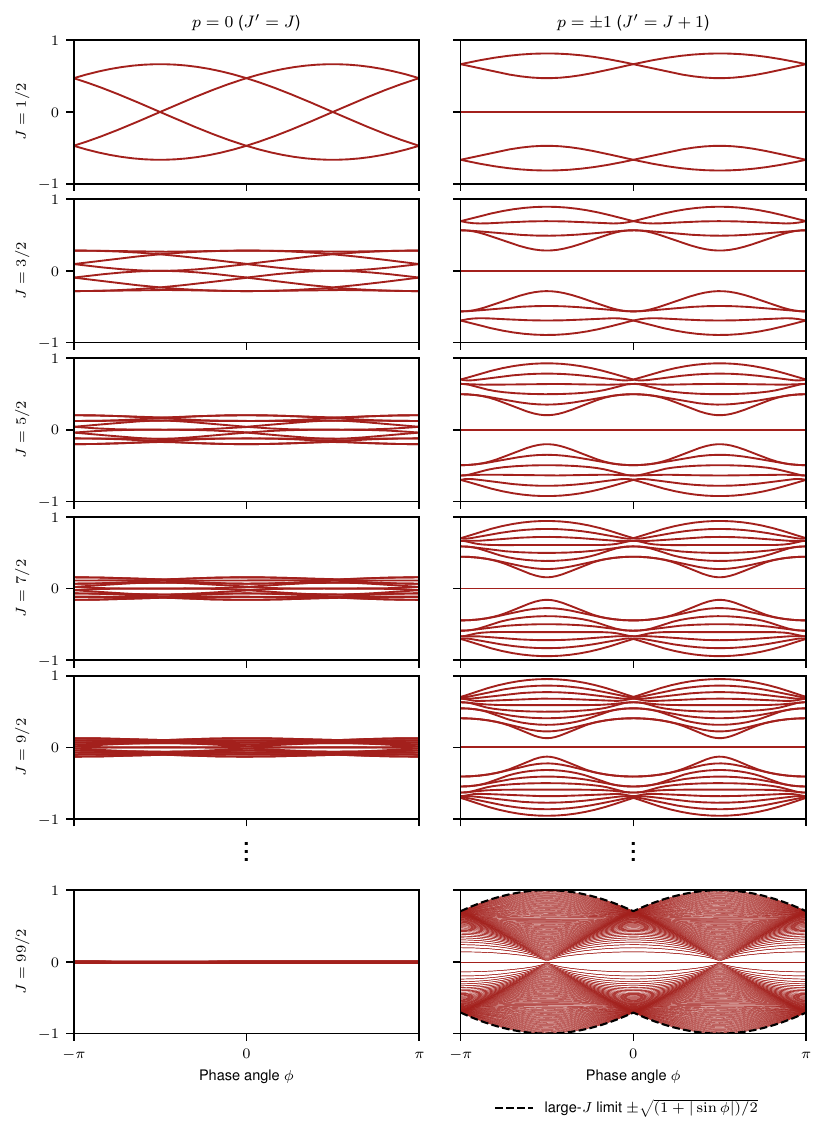}
    \caption{\flb{Atlas over angular eigenvalue variation for coupling matrix \eref{eq:big}} As $J$ increases the $p=0$ spectra collapse onto a line, while the $p=\pm 1$ spectra attains an outer envelope described by $\pm\sqrt{(1+|\sin\phi|)/2}$.}\label{fig:sup:moreJ}
\end{figure}

\normalsize
Using the explicitly expressed coupling matrix \eref{eq:big} we can compute the eigenvalue spectrum as a function of the phase angle $\phi$. Supplementary \fref{fig:sup:moreJ} shows the development in eigenvalue spectra as $J$ increases
\newpage
\section{Beyond a two-level picture}\label{sec:sup:3level}
\begin{figure}[b!]
\includegraphics[width=0.43\linewidth]{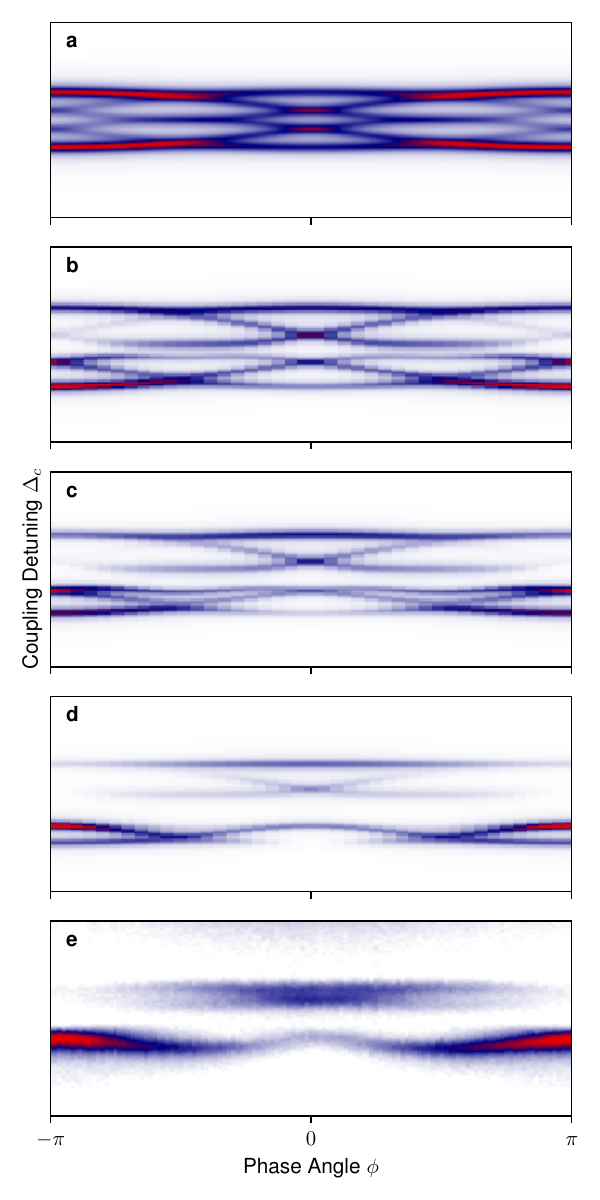}
    \caption{\flb{Beyond a dressed two-level system} Effect of coupling off-resonantly to a third level. \lb{a} Simulated  spectrogram for an isolated \typ{3}{0}-transition nominally occurring between the two levels $49D_{3/2}$ and $50P_{3/2}$. \lb{b},\lb{c} spectrograms including coupling to the  $49D_{5/2}$ level placed, respectively,  \qty{350}{\mega\hertz} and \qty{200}{\mega\hertz} above the $49D_{3/2}$ -state. \lb{d} spectrogram computed with the actual separation of \qty{98.7}{\mega\hertz}. \lb{e}  Experimentally observed spectrogram when resonantly dressing the $49D_{3/2} \leftrightarrow 50 P_{3/2}$ transition. In doing so  the  $49D_{5/2} \leftrightarrow 50P_{3/2}$ transition is off-resonantly dressed and correspondence with \lb{d} rather than \lb{a} is found.}\label{fig:sup:three-rydberg}
\end{figure}
Our experimental observations of the $J^p = \frac{3}{2}^0$ transition  resemble neither the analytic nor the simulated spectra obtained in a two-level model. The Rydberg states we address with an RF field in the experiment are the $49D_{3/2}$ and $50P_{3/2}$ levels of $\rm^{87}Rb$. In addition a $49D_{5/2}$ level is located 98.7~MHz above the $49D_{3/2}$ state as calculated with Alkali Rydberg Calculator (ARC) \cite{Sibalic2017}. It is the presence of this state that distorts the ideal $J^p = \frac{3}{2}^0$ spectrum. Supplementary \Fref{fig:sup:three-rydberg}a-d shows density matrix simulations of the spectrogram deformation as a $D_{5/2}$ state is moved into its physical location (Supplementary \fref{fig:sup:three-rydberg}d) from far away (Supplementary \fref{fig:sup:three-rydberg}a), where an ideal $J^p = \frac{3}{2}^0$ spectrogram is obtained. The spectrogram Supplementary \fref{fig:sup:three-rydberg}d, which includes the effect of the $D_{5/2}$ level clearly provides a better description of the experimentally obtained spectrogram Supplementary \fref{fig:sup:three-rydberg}e

\begin{figure}
\includegraphics[width=0.8\linewidth]{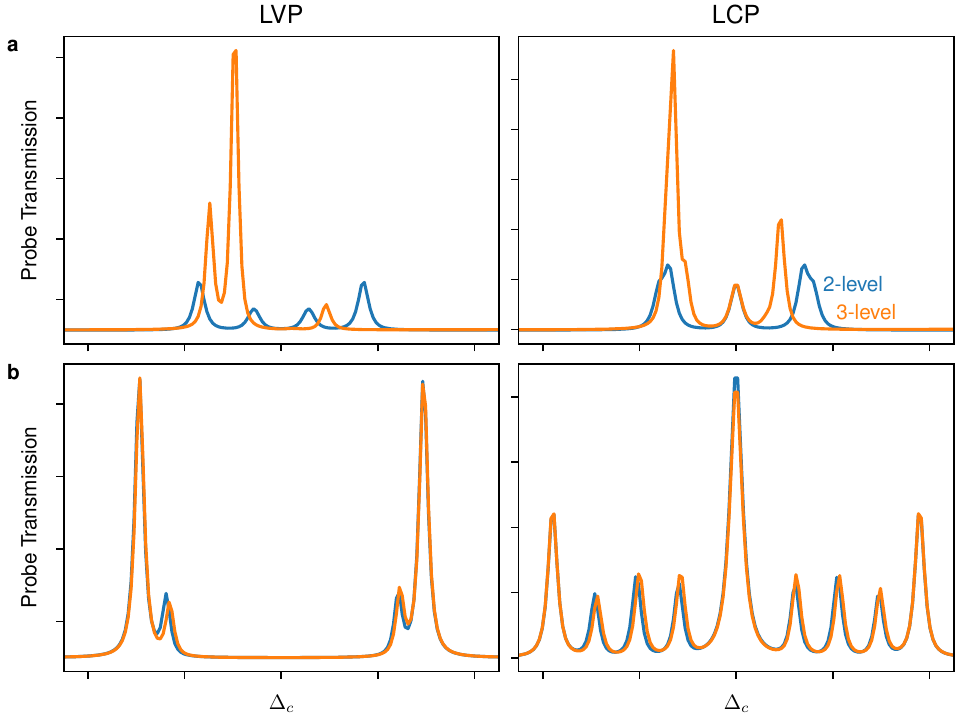}
\caption{\flb{Simulated EIT spectra for two- and three-level systems} \textbf{a} $J^p = \frac{3}{2}^0$ $\updownarrow^{49D_{3/2}}_{50P_{3/2}}$ system with (orange) and without (blue) perturbation from the $49D_{5/2}$ level present. \textbf{b} $J^p = \frac{3}{2}^+$ $\updownarrow^{49D_{5/2}}_{50P_{3/2}}$ with (orange) and without (blue) perturbation from the $49D_{3/2}$ level present. For both \textbf{a} and \textbf{b}, spectra with LVP and LCP RF dressing fields are shown.}\label{fig:sup:D5-three-rydberg}
\end{figure}

The $J^p = \frac{3}{2}^+$ transition studied in this work and used to demonstrate RF polarimetry is physically embodied as the transition between the $50P_{3/2}$ and $49D_{5/2}$ levels of $\rm^{87}$Rb. This raises a question about the extent to which $49D_{3/2}$ will affect the idealized $J^p = \frac{3}{2}^+$ two-level scenario. Compared to $J^p = \frac{3}{2}^0$ $\updownarrow^{49D_{3/2}}_{50P_{3/2}}$ perturbed by $49D_{5/2}$, the effect on $J^p = \frac{3}{2}^+$ $\updownarrow^{49D_{5/2}}_{50P_{3/2}}$ from the $49D_{3/2}$ level is very small. Supplementary \fref{fig:sup:D5-three-rydberg} shows simulated EIT spectra for the two cases demonstrating that in the latter case, the difference between a two- and three-level description is close to indistinguishable.

\clearpage

\section{Matrix elements of \texorpdfstring{$r_q$}{r\_q}}
\label{sec:sup:rq_matrix}
We want to calculate the dipole matrix element between an initial quantum state $\ket{i}$ and a final state $\ket{f}$ in the $J=L+S$ coupled basis under the assumptions:
\begin{enumerate}
    \item $S=S'=1/2$.
    \item For the dipole matrix element to be non-zero, $L$ and $L'$ must differ by 1 (Laporte's rule). Without loss of generality we will assume $L'=L+1$. This implies that for allowed transitions, we either have $J'=J$ or $J'=J+1$.
    \item The wavefunction for the Rydberg level corresponding to a principal quantum number $n$ and orbital angular momentum $L$ is described by a radial component $R_{nL}(r)$.
    \end{enumerate}
    
Using the Wigner-Eckart theorem [cf.~Ref.~\onlinecite{Sobelman1992}, (4.120) and (4.175), and Ref.~\onlinecite{Thompson1994b}, (8.46)]  we obtain

\begin{align}
\bra{n'L'S'J'm_J'} r_q \ket{nLSJm_J}
&= (-1)^{J'-m_J'}
\begin{pmatrix}
 J' & 1 & J \\
 -m_J' & q & m_J
\end{pmatrix}
\ReducedMat{n'L'S'J'}{r}{nLSJ} \nonumber \\[0.5em]
&= (-1)^{J'-m_J'}
\begin{pmatrix}
 J' & 1 & J \\
 -m_J' & q & m_J
\end{pmatrix}
(-1)^{L'+S'+J+1}
\nonumber \\
&\quad\times\sqrt{(2J+1)(2J'+1)} 
\begin{Bmatrix}
 L' & J' & S' \\
 J  & L  & 1
\end{Bmatrix}
\ReducedMat{n'L'}{r}{nL}
 \nonumber \\[0.5em]
&= (-1)^{J'-m_J'}
\begin{pmatrix}
 J' & 1 & J \\
 -m_J' & q & m_J
\end{pmatrix}
(-1)^{L'+S'+J+1}
\nonumber \\
&\quad\times\sqrt{(2J+1)(2J'+1)} 
\begin{Bmatrix}
 L' & J' & S' \\
 J  & L  & 1
\end{Bmatrix}
{(-1)^{L'}\sqrt{2L+1}\sqrt{2L'+1}\underbrace{\begin{pmatrix}
 1 & L'=L+1 & L \\
 0 & 0 & 0
\end{pmatrix}}_{(-1)^{L+1}\sqrt{\frac{L+1}{(2L+1)(2L+3)}}}{\mathcal{R}_{n'L',nL}}}
 \nonumber \\[0.5em]
&= (-1)^{J'+J+L-m_J'+\frac12}
\begin{pmatrix}
 J' & 1 & J \\
 -m_J' & q & m_J
\end{pmatrix}
 \nonumber \\
&\quad\times
\underbrace{\sqrt{(2J+1)(2J'+1)(L+1)}\begin{Bmatrix}
 L+1 & J' & \tfrac12 \\
 J   & L  & 1
\end{Bmatrix}}_{\Pi}
\mathcal{R}_{n'L+1,nL},
\label{eq:dipole}
\end{align}
where
\begin{equation}
   {\mathcal{R}_{n'L',nL}}\equiv{\int_0^\infty R_{n'L'}(r)R_{nL}(r)r^3dr},
\end{equation}
is the radial matrix element for the $i\rightarrow f$ transition, which has no $J$ or $m_J$ dependence.

In \eref{eq:dipole} the 3-$j$ symbol expresses the relative transition amplitudes across the $m_J$-manifold. Meanwhile, the product $\Pi$ in the last line of \eref{eq:dipole} simplifies as follows.

    With the assumption $L'=L+1$ there are three possibilities for dipole-allowed transitions:
    \begin{description}
     \item [p=+1] $J'=J+1=L'+1/2=L+3/2$ so that for \eref{eq:dipole} 
        \begin{enumerate}
     \item 
        the six-$j$ symbol reads [using Ref. \nocite{Varshalovich1988}\onlinecite{Varshalovich1988} 9.4.2(2) and  9.5.2 (4)]\\ $\begin{Bmatrix}
		L+1& L+\frac{3}{2} & \frac{1}{2}\nonumber\\
		L+\frac{1}{2}& L & 1
	\end{Bmatrix}=\begin{Bmatrix}
		L& 1 & L+1\nonumber\\
		L+\frac{3}{2}& \frac{1}{2} & L+\frac{1}{2}
	\end{Bmatrix}=\frac{-1}{\sqrt{(2L+3)(2L+2)}}=\frac{-1}{\sqrt{(2J+2)(2J+1)}}$.
    \item $\sqrt{(2J+1)(2J'+1)(L+1)}=\sqrt{(2J+1)(2J+3)(J+\frac{1}{2})}=\sqrt{\frac{1}{2}(2J+1)^2(2J+3)}$.
    \item the product of these is $-\sqrt{\frac{(2J+1)(2J+3)}{2(2J+2)}}=-\sqrt{(2J+3)}\sqrt{\frac{(2J+1)}{2(2J+2)}}$
    \end{enumerate}
        \item [p=0] $J'=J=L+1/2$ so that 
        for \eref{eq:dipole}
    \begin{enumerate}
            \item 
            the six-$j$ symbol reads [using Ref. \nocite{Varshalovich1988}\onlinecite{Varshalovich1988} 9.4.2(2) and 9.5.2(4)]\newline $\begin{Bmatrix}
    		L+1& L+\frac{1}{2} & \frac{1}{2}\nonumber\\
    		L+\frac{1}{2}& L & 1
    	\end{Bmatrix}=\begin{Bmatrix}
    		\frac{1}{2}& L+\frac{1}{2} & L+1 \nonumber\\
    		1& L & L+\frac{1}{2}
    	\end{Bmatrix}=\sqrt{
        \frac{2}{(2L + 1)(2L + 2)^2(2L + 3)}
        }=\sqrt{
        \frac{1}{J(2J + 1)^2(2J + 2)}
        }$
        \item $\sqrt{(2J+1)(2J'+1)(L+1)}=\sqrt{(2J+1)^2(J+\frac{1}{2})}=\sqrt{\frac{(2J+1)^3}{2}}$
        \item the product of these is $\sqrt{\frac{(2J+1)}{2J(2J+2)}}=\frac{1}{\sqrt{J}}\sqrt{\frac{(2J+1)}{2(2J+2)}}$
    \end{enumerate}

 \item[p=-1] $J'=J+1=L'-1/2=L+1/2$ so that for \eref{eq:dipole} 
        \begin{enumerate}
     \item 
        the six-$j$ symbol reads [using Ref. \nocite{Varshalovich1988}\onlinecite{Varshalovich1988} 9.4.2(2) and  9.5.2 (5)]\\ $\begin{Bmatrix}
		L+1& L+\frac{1}{2} & \frac{1}{2}\nonumber\\
		L-\frac{1}{2}& L & 1
	\end{Bmatrix}=\begin{Bmatrix}
		\frac{1}{2}& L+\frac{1}{2} & L+1\nonumber\\
		1& L & L-\frac{1}{2}
	\end{Bmatrix}=\frac{1}{\sqrt{(2L+2)(2L+1)}}=\frac{1}{\sqrt{(2J+3)(2J+2)}}$.
    \item $\sqrt{(2J+1)(2J'+1)(L+1)}=\sqrt{(2J+1)(2J+3)(J+\frac{3}{2})}=\sqrt{\frac{(2J+1)(2J+3)^2}{2}}$.
    \item the product of these is $\sqrt{\frac{(2J+1)(2J+3)}{2(2J+2)}}=\sqrt{(2J+3)}\sqrt{\frac{(2J+1)}{2(2J+2)}}$.
    \end{enumerate}
     \end{description}
   Altogether we can cast \eref{eq:dipole} as
   \begin{subequations}
   \begin{align}
\bra{n'L'S'J'm_J'} r_q \ket{nLSJm_J}
&= (-1)^{J'+J+L-m_J'+\frac12-(p+p^2)/2}{\scriptstyle
\sqrt{\frac{(2J+1)}{4(J+1)}}\sqrt{\frac{(2J+3)^{|p|}}{J^{(1-|p|)}}}
}
\begin{pmatrix}
 J' & 1 & J \\
 -m_J' & q & m_J
\end{pmatrix}\mathcal{R}_{n'L+1,nL},
\label{eq:dipole2}\nonumber \\&=(-1)^{J+m'_J+|p|}
\sqrt{\frac{2J+1}{4(J+1)}}
\sqrt{\frac{(2J+3)^{|p|}}{J^{1-|p|}}}
\begin{pmatrix}
J & 1 & J+|p|\\
m_J & q & -m'_J
\end{pmatrix}
\mathcal R_{n'L+1,nL}.
\end{align}
and, using $r_q^\dagger=(-1)^q r_{-q}$ and the reality of the 3$j$ and 6$j$ symbols and of the radial integral,
\begin{align}
\langle nLSJm_J|\,r_q\,|n'L'S'J'm'_{J}\rangle
&=(-1)^q\langle n'L'S'J'm'_{J}|\,r_{-q}\,|nLSJm_J\rangle\nonumber\\
&=(-1)^{J+m'_{J}+|p|+q}\sqrt{\frac{2J+1}{4(J+1)}}\sqrt{\frac{(2J+3)^{|p|}}{J^{1-|p|}}}
\begin{pmatrix}J&1&J+|p|\\ m_J&-q&-m'_{J}\end{pmatrix}\mathcal R_{n'L+1,nL}. \label{eq:3b}
\end{align}
\end{subequations}
Using an ordered basis,\[
\bigl\{
\overbrace{\ket{J,-J}}^{1},\ldots,\overbrace{\ket{J,J}}^{2J+1},
\overbrace{\ket{J',-J'}}^{2J+2},\ldots,\overbrace{\ket{J',J'}}^{4J+2|p|+2}
\bigr\},
\] we can arrange the elements defined by \eref{eq:dipole2} and \eref{eq:3b} in
a matrix $\bm T^{(Jp,q)}$ of dimension $(2J+2J'+2)\times(2J+2J'+2)$ with two non-zero off-diagonal blocks. Since the columns of this matrix $j=1...(2J+1)$ correspond to the kets $\ket{J,m_J=-J}...\ket{J,m_J=J}$, while the rows $i=2(J+1)...4J+2|p|+2$ correspond to the bras $\bra{J'=J+|p|,m'_{J}=-(J+|p|)}...\bra{J'=J+|p|,m'_{J}=J+|p|}$, an entry $ij$ in the lower non-zero block can be expressed as
\begin{subequations}

\begin{align}
T^{(Jp,q)}_{ij}
=(-1)^{i+1}
\sqrt{\frac{2J+1}{4(J+1)}}
\sqrt{\frac{(2J+3)^{|p|}}{J^{1-|p|}}}
\begin{pmatrix}
J & 1 & J+|p|\\
j-1-J & q & -(i-3J-2-|p|)
\end{pmatrix}
\mathcal R_{n'L+1,nL},
\end{align}
while for the upper non-zero block, where the rows $i=1\ldots(2J+1)$ correspond to the bras $\langle J,m_J=-J|\ldots\langle J,m_J=J|$ and the columns $j=2(J+1)\ldots4J+2|p|+2$ correspond to the kets $|J'=J+|p|,m'_{J'}=-(J+|p|)\rangle\ldots|J'=J+|p|,m'_{J'}=J+|p|\rangle$, we have
\begin{align}
T^{(Jp,q)}_{ij}
=(-1)^{j+1+q}\sqrt{\frac{2J+1}{4(J+1)}}\sqrt{\frac{(2J+3)^{|p|}}{J^{1-|p|}}}
\begin{pmatrix}J&1&J+|p|\\ i-1-J&-q&-(j-3J-2-|p|)\end{pmatrix}\mathcal R_{n'L+1,nL}.
\end{align}
\end{subequations}
Assuming $q=\pm1$ a general element of $\bm T^{(Jp,q)}$ can be expressed as
\begin{equation}
T_{ij}^{(Jp,q)}=\sqrt{\tfrac{2J+1}{4(J+1)}}\!\sqrt{\tfrac{(2J+3)^{|p|}}{J^{1-|p|}}}
\bigl[\,(-1)^{i+1}\!\bigl(\begin{smallmatrix}J & 1 & J+|p|\\ j{-}1{-}J & q & 3J{+}2{+}|p|{-}i\end{smallmatrix}\bigr)
+(-1)^{j}\!\bigl(\begin{smallmatrix}J&1&J+|p|\\ i{-}1{-}J&-q&3J{+}2{+}|p|{-}j\end{smallmatrix}\bigr)\,\bigr]\mathcal R_{n'L+1,nL}.
\label{eq:S5}
\end{equation}
\Eref{eq:Efield} gives the RF field in the spherical basis $\hat{\boldsymbol e}_{\pm1}=\mp(\hat{\boldsymbol x}\pm i\hat{\boldsymbol y})/\sqrt2$ as
\begin{equation*}
\boldsymbol E(t)=\frac{E_0}{\sqrt2}\,\mathrm{Re}\Bigl\{\boldsymbol\varepsilon\,e^{-i(\omega t-\phi/2)}\Bigr\},\qquad
\boldsymbol\varepsilon=\sum_{q=\pm1}\varepsilon_q\hat{\boldsymbol e}_q,\qquad
\varepsilon_q=-q\Bigl(\cos\frac{\phi}{2}-q\sin\frac{\phi}{2}\Bigr).
\end{equation*}
With $\boldsymbol r\cdot\hat{\boldsymbol e}_q=r_q$ and $\boldsymbol r\cdot\hat{\boldsymbol e}_q^{\,*}=(-1)^q r_{-q}=r_q^\dagger$, the interaction $-\boldsymbol d\cdot\boldsymbol E=e\,\boldsymbol r\cdot\boldsymbol E$ reads
\begin{equation}
-\boldsymbol d\cdot\boldsymbol E(t)=\frac{eE_0}{2\sqrt2}\sum_{q=\pm1}\varepsilon_q\Bigl[r_q\,e^{-i(\omega t-\phi/2)}+r_q^\dagger\,e^{i(\omega t-\phi/2)}\Bigr].
\label{eq:S6}
\end{equation}
In a frame rotating at $\omega$, the rotating-wave approximation retains from $r_q$ only its upper block $\langle J|r_q|J'\rangle$ and from $r_q^\dagger=(-1)^q r_{-q}$ only its lower block $(-1)^q\langle J'|r_{-q}|J\rangle$. Both are contained in equation~\eqref{eq:S5}: the upper block is its second term, evaluated at $q$, and the lower block is its first term, evaluated at $-q$ and multiplied by $(-1)^q$, which turns $(-1)^{i+1}$ into $(-1)^{i}$. Absorbing the common phase $e^{i\phi/2}$ into the states $|J',m'_{J'}\rangle$, we obtain $H=\frac{eE_0\mathcal R_{n'L+1,nL}}{2\sqrt2}\boldsymbol M^{(Jp)}$ with
\begin{equation}
\begin{split}
M^{(Jp)}_{ij}=\sqrt{\frac{2J+1}{4(J+1)}}\sqrt{\frac{(2J+3)^{|p|}}{J^{1-|p|}}}\sum_{q=\pm1}\varepsilon_q
\biggl[&(-1)^{i}\begin{pmatrix}J&1&J+|p|\\ j-1-J&-q&3J+2+|p|-i\end{pmatrix}\\
&+(-1)^{j}\begin{pmatrix}J&1&J+|p|\\ i-1-J&-q&3J+2+|p|-j\end{pmatrix}\biggr],
\end{split}
\label{eq:S7}
\end{equation}
which is manifestly symmetric in $i$ and $j$, as required for a Hermitian coupling with real entries. Reversing the signs in the bottom row of each 3$j$ symbol multiplies it by $(-1)^{2J+1+|p|}$; with the selection rule $i=j+2J+1+|p|-q$ of the first 3$j$ symbol, the first term becomes
\begin{equation*}
\begin{split}
&\varepsilon_q(-1)^{i}(-1)^{2J+1+|p|}\begin{pmatrix}J&1&J+|p|\\ J+1-j&q&i-3J-|p|-2\end{pmatrix}\\
&\qquad=q\,(-1)^{j}\Bigl(\cos\frac{\phi}{2}-q\sin\frac{\phi}{2}\Bigr)\begin{pmatrix}J&1&J+|p|\\ J+1-j&q&i-3J-|p|-2\end{pmatrix},
\end{split}
\end{equation*}
since $(-1)^{i+2J+1+|p|}=(-1)^{j-q}=-(-1)^{j}$ for $q=\pm1$, and likewise for the second term with $i\leftrightarrow j$. This is equation~(\ref{eq:big}) of the main text. The common prefactor $eE_0\mathcal R_{n'L+1,nL}/(2\sqrt2)$ sets the overall scale of the eigenvalues and cancels in all ratios of splittings.

\clearpage\newpage
\section{Eigenvalue envelopes for \typ{3}{\pm} transitions}\label{sec:sup:envelope}
     The envelopes bounding the positive and negative eigenvalues can be expressed analytically. For \typ{3}{\pm} transitions the outer envelopes are described by
 \begin{subequations}    \begin{equation}\epsilon_{\rm o\pm}(\phi)=\pm
\frac{1}{5}\sqrt{
10 + 3\lvert \sin \phi\rvert
+ \sqrt{33\sin^2 \phi + 12\lvert \sin \phi\rvert + 4}
},\label{eq:env1}
\end{equation}
whereas the inner envelopes are given by
\begin{equation}\epsilon_{\rm i\pm}(\phi)=\pm
\frac{1}{5}\sqrt{
10 - 3\lvert \sin \phi\rvert
- \sqrt{33\sin^2 \phi - 12\lvert \sin \phi\rvert + 4}
}.\label{eq:env2}
\end{equation}
 \end{subequations} 
Supplementary \Fref{fig:env}a shows the envelope functions defined by \eref{eq:env1} and \eref{eq:env2}. While these expressions are exact, they are cumbersome and may conveniently be approximated as 
 \begin{subequations}
    \begin{figure}[b!] \includegraphics[width=\linewidth]{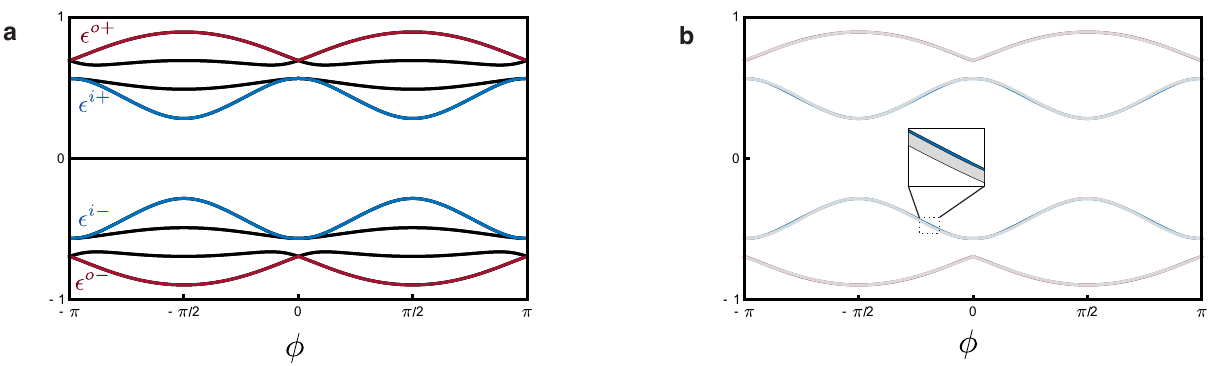}
    \caption{\flb{Eigenvalue envelopes} \lb{a}  Exact envelopes $\epsilon_{\rm o\pm}$ and $\epsilon_{\rm i\pm}$ as given by \eref{eq:env1} (red line) and \eref{eq:env2} (blue line), respectively.  \lb{b} Approximate envelopes (grey line) as given by  \eref{eq:enva1} and \eref{eq:enva2} overlaid on the exact envelopes (red and blue lines) from which they are close to indistinguishable. To show that there is in fact a small discrepancy,  a zoom-in is included at the angle where the maximum $\sim1\%$ deviation occurs for $\bar{\epsilon}_{\rm i}$. \label{fig:env}}
\end{figure}

\begin{equation}
\bar{\epsilon}_{\rm o\pm}(\phi)=\pm\frac{2\sqrt{3}}{5}
\pm \frac{2(\sqrt{5}-\sqrt{3})}{5}\,|\sin \phi|,\label{eq:enva1}\end{equation}
 for the outer envelopes, and
\begin{equation}
\bar{\epsilon}_{\rm i\pm}(\phi)=\pm\frac{\sqrt{2}}{5}\left(2 - \sin^2 \phi\right),\label{eq:enva2}
\end{equation}
\end{subequations}
for the inner. Supplementary  \Fref{fig:env}b shows that these approximations are extremely good. The maximum deviations across the domain of the functions are $\lesssim 0.1\%$ for $\bar{e}_{\rm o\pm}$ and $\lesssim 1\%$ for $\bar{e}_{\rm i\pm}$.

Analogous to \eref{eq:ratio} we define
  \begin{equation}
R=\frac{\bar{\epsilon}_{\rm i+}(\phi)-\bar{\epsilon}_{\rm i-}(\phi)}{\bar{\epsilon}_{\rm o+}(\phi)-\bar{\epsilon}_{\rm o-}(\phi)},\label{eq:ratio2}
  \end{equation}
  which is a quantity that can be determined from an experimentally acquired spectrum.
\begin{figure}[b!] \includegraphics[width=0.5\linewidth]{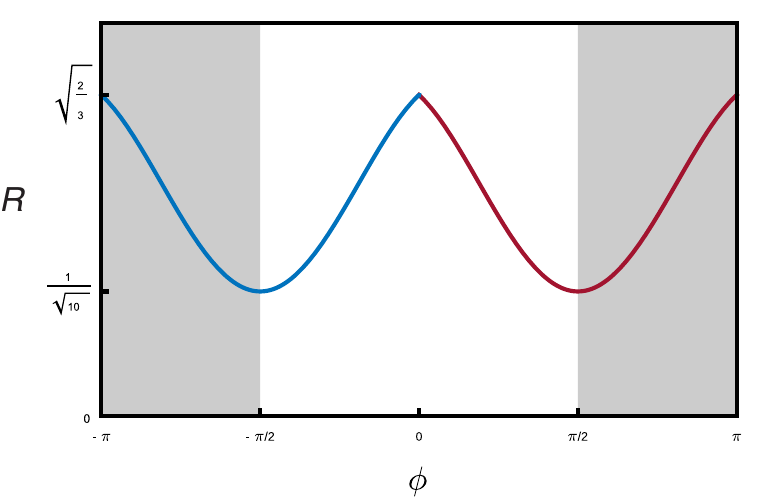}
    \caption{\flb{Eigenvalue envelope ratio $R$ for \typ{3}{\pm} transitions} The variation of $R$ with the phase angle $\phi$ of the RF dressing field as given by \eref{eq:ratio2}. When inverting a given value of $R$ to possible values of $\phi$, auxiliary information from the prominence of the central EIT peak can be
    used to restrict solutions to lie either inside or outside the grey shaded interval. In both cases one is left with a ``blue'' and a ``red'' solution, because the phase angles $\pm\phi$ give rise to identical spectra. 
    Carrying out the operation \fref{fig:45}a $\rightarrow$ \fref{fig:45}b does not affect $R(\phi)$, but it will shift the unshaded region to the interval $[0,\pi]$, and it therefore becomes possible to discriminate between the red and blue candidates. \label{fig:envrat}}
\end{figure}
$R(\phi)$ is plotted in Supplementary \fref{fig:envrat}, from which it is clear that a measured value for $R$ is generally compatible with four possible phase angles. $R=\sqrt{2/3}\approx0.82$ and $R=\sqrt{1/10}\approx0.32$ are special cases of twofold ambiguity corresponding to LHP/LVP and RCP/LCP, respectively (cf. \fref{fig:setup}b).

For a given value of $R$ the four possible phase angles can be found by solving the quadratic equation
  \begin{equation}
      2\sqrt{3}
+ 2(\sqrt{5}-\sqrt{3})\,|\sin \phi|=R^{-1}\sqrt{2}\left(2 - \sin^2 \phi\right),
  \end{equation}
    and they are found to be
\begin{equation}
    \phi=\pm\phi_{\rm P}, \pm(\pi-\phi_{\rm P}),
\end{equation}
where
    \begin{equation}
    \phi_{\rm P}
=\arcsin\!\left[
\sqrt{2-\sqrt{6}\,R+(4-\sqrt{15})R^2}
-({\sqrt{5/2}-\sqrt{3/2}})\,R
\right],
    \end{equation}
    is the principal value solution.

We note that, unlike a $J^0$-transition, the dressed state spectrum for a \typ{3}{\pm} transition includes a central peak. In our optical EIT probing of the dressed-state energies of $\updownarrow^{J'}_J=\updownarrow^{5/2}_{3/2}$ using the configuration of \fref{fig:setup}a, the prominence of this central peak displays a $\phi$-dependence: it vanishes identically for $\phi=\pm \pi$, while attaining its maximum at $\phi=0$. As discussed in the manuscript, this breaks a symmetry for the spectrograms that we record for the $\updownarrow^{J'}_J=\updownarrow^{5/2}_{3/2}$-transition (see \fref{fig:zoo}b). Taking into account the auxiliary information provided by the prominence of the central peak in a given spectrum, we can ascertain whether $\phi$ lies inside or outside the interval $[-\frac{\pi}{2},\frac{\pi}{2}]$; Supplementary \fref{fig:envrat} shows the region outside this interval as shaded. 
\clearpage
\newpage 
\section{Spectral inversion procedure}\label{sec:sup:inv}

This note describes how the phase angles shown in \figref{fig:inversion} of the main text are obtained from the
measured EIT spectra of the $59S_{1/2}\leftrightarrow59P_{1/2}$ ($J^p=\tfrac12^{0}$, \figref{fig:inversion}a) and
$50P_{3/2}\leftrightarrow49D_{5/2}$ ($J^p=\tfrac32^{+}$, \figref{fig:inversion}b) transitions. Each data set
comprises 121 spectra recorded at set phases spaced by $3^\circ$ over a full period. The analysis is guided by three principles.
(i)~Every spectrum is analysed on its own; no information is taken from spectra at other set
phases, with the single exception of the line centre (see below), which carries no information on
$\phi$. (ii)~The phase angle is inferred from line \emph{positions} only. (iii)~The set phases are
used exclusively to score the result after the analysis has been completed.

\subsection*{Information contained in a single spectrum.}
The line positions are given by the eigenvalues of the coupling matrix $\mathbf{M}^{(Jp)}(\phi)$,
\eref{eq:big}. For both transitions these eigenvalues depend on $\phi$ only through $|\sin\phi|$,
i.e.\ on the ellipticity of the RF field. A single spectrum therefore determines
$\phi_0=\arcsin|\sin\phi|\in[0,\pi/2]$ and hence the fourfold set of candidates
$\{\pm\phi_0,\;\pm(\pi-\phi_0)\}$. Throughout, angles are given in the convention of
\eref{eq:phi}.

\subsection*{Preprocessing}
The EIT spectrum without RF defines the detuning axis, with the resonance position providing the zero, and the line-width of a Lorentzian giving the length scale for the rest of the procedure. This is followed by an extra line centring step: any small residual error in the zero position is then estimated and removed by finding any strong peaks in the dataset near the bare resonance zero. The axis is then shifted so that their average position is the new zero.

\subsection*{Line detection.}
Each spectrum is smoothed with a Gaussian of standard deviation 0.1 linewidths.
Its own baseline is defined as the lower envelope of the smoothed spectrum (running minimum over
6 linewidths, smoothed with a Gaussian of 3 linewidths), and every local maximum is characterised by its
height above this baseline.  Maxima closer together than a
single linewidth are counted as a single line. A line is retained if its height exceeds a
fraction $\eta$ of a reference height in the same spectrum: the tallest line for
$J^p=\tfrac12^0$, and the tallest non-central line for $J^p=\tfrac32^+$, where the central line can
be several times stronger than all others. Lines within $\pm 1$ line widths of zero detuning are classified as
central and are not used in the envelope ratio.

\subsection*{Mirror pairs and envelope ratio.}
Since the eigenvalues occur in pairs $\pm\lambda_k$, the non-central lines must come in pairs that
are mirror symmetric about $c$. A line at $\Lambda_-<c$ and a line at $\Lambda_+>c$ form a candidate
pair if $|\Lambda_+-(2c-\Lambda_-)|<\delta$. Candidate pairs are accepted in order of the height of
their weaker member, so that among several candidates within the tolerance the tallest partner is
chosen, and no line is used twice. At most two pairs ($J^p=\tfrac12^0$) or four pairs
($J^p=\tfrac32^+$) are accepted. The positions of the accepted lines
are refined by a least-squares fit of a Lorentzian plus a linear background within $\pm0.8$ linewidths,
with the centre constrained to $\pm0.25$ linewidths of the detected maximum. The outermost and innermost
accepted pairs define the envelope ratio of \eref{eq:ratio},
\begin{equation*}
  R=\frac{\Lambda_{\mathrm{i}+}-\Lambda_{\mathrm{i}-}}{\Lambda_{\mathrm{o}+}-\Lambda_{\mathrm{o}-}} .
\end{equation*}
If only one pair is resolved, its inner and outer envelopes coincide and $R=1$; for
$J^p=\tfrac12^0$ a single pair together with a central line instead indicates a collapsed inner
pair and gives $R=0$. A spectrum without any mirror pair is reported as a failure.
The parameters are summarised in Supplementary Table~\ref{tab:params}.

\begin{table}[h]
\centering
\caption{Analysis parameters (detuning in the linewidths).}
\label{tab:params}
\begin{tabular}{lcc}
\toprule
 & $J^p=\tfrac12^0$ & $J^p=\tfrac32^+$\\
\midrule
detection threshold $\eta$                   & 30\,\% of tallest line & 20\,\% of tallest non-central line\\
mirror-pair tolerance $\delta$               & $\pm20$ & $\pm20$ \\
maximum number of pairs                      & 2     & 4     \\
\bottomrule
\end{tabular}
\end{table}

\subsection*{Inversion.}
For $J^p=\tfrac12^0$, $\phi_0$ follows from \eref{eq:Rinv}. For $J^p=\tfrac32^+$, $\phi_0$ follows
from the closed-form expression \eref{eq:phitilde}, based on the approximate envelopes of equations
(\ref{eq:enva1},b). Over the full range of $R$ it agrees with the exact inversion, obtained numerically
from the eigenvalues of equation~(\ref{eq:big}), to within $1^\circ$ (mean difference $0.4^\circ$ for the
present data). 

\subsection*{Branch decision for \texorpdfstring{$J^p=\tfrac32^+$}{Jp=3/2+} (\figref{fig:inversion}b).}
In the probe configuration of \fref{fig:setup}a the central EIT peak is strong when the major axis of the
polarization ellipse is horizontal ($|\phi|<\pi/2$) and vanishes for linear vertical polarization
($|\phi|=\pi$). We quantify this within each spectrum by $\kappa$, the height of the central line
divided by the height of the tallest non-central line. For $\kappa\ge3$ we set $|\phi|=\phi_0$,
otherwise $|\phi|=\pi-\phi_0$, which leaves the two candidates $\pm|\phi|$ of opposite helicity
shown in \figref{fig:inversion}b. This is the only step that uses line heights. $\kappa$ varies smoothly with
$\phi$, from $\approx0.05$ near $|\phi|=\pi$ to $4$--$6$ for $|\phi|<\pi/2$, but its value at the
boundary $|\phi|=\pi/2$ depends on the optical fields and cannot be derived from the line
positions; the threshold $\kappa=3$ was therefore set once from the measured dependence of
$\kappa$ on the set phase and constitutes a characterization of the optical configuration. With this threshold,
8 spectra are assigned to the wrong branch, all close to circular polarization, where the two
branches converge. 

\subsection*{Results.}
The parameters in Supplementary Table~\ref{tab:params} were chosen by inspecting the spectra and
the inversion results for a subset of the data, and the same parameters are applied to all
spectra. Deviations are evaluated as the distance between the set phase and the nearest candidate.
Supplementary Table~\ref{tab:results} summarises the results for all 121 spectra of each
transition.

\begin{table}[h]
\centering
\caption{Accuracy of the inversion: number of spectra analysed, median and r.m.s.\ deviation of
the inferred from the set phase angle, and number of deviations exceeding $15^\circ$.}
\label{tab:results}
\begin{tabular}{lcccc}
\toprule
Transition & Analysed & Median & r.m.s. & $>15^\circ$\\
\midrule
$J^p=\tfrac12^0$ (\figref{fig:inversion}a)  & 120/121 & $2.6^\circ$ & $4.5^\circ$  & 1\\
$J^p=\tfrac32^+$ (\figref{fig:inversion}b)  & 121/121 & $3.1^\circ$ & $10.5^\circ$ & 11\\
\bottomrule
\end{tabular}
\end{table}
\subsection*{Limitations.}
For $J^p=\tfrac12^0$ the spectra that cannot be analysed lies next to circular polarization,
where an outer line is double-peaked or too weak to be detected; it is not shown in \figref{fig:inversion}a.
The remaining deviations of up to $\approx10^\circ$ occur near linear and circular polarization,
where lines merge and $\phi_0$ is least well determined. For $J^p=\tfrac32^+$, we again see structure in the deviations. Especially for $\phi$ just below $90^\circ$, one member of the innermost pair is too weak to be detected in the spectra, so that the envelope ratio is formed with the second pair. The effect appears for both
helicities and reflects the dependence of line strengths on the probing configuration (cf. \figref{fig:45}); it is removed by averaging over the orientation $\alpha$ of the optical axis, as in the
$\alpha$-scan protocol (cf. the improvement in \figref{fig:inversion_alpha}). An isolated outlier near $\phi=0^\circ$ arises
when a shoulder of the strong central line pairs with a weak feature. Excluding a wider zone around
the central line removes this outlier but causes extra failures in spectra with a weak central
line. The remaining deviations above $15^\circ$
are spectra close to linear polarization, and one spectrum assigned to the wrong branch close to circular
polarization.

\clearpage\newpage
\section{SOP inversion from \texorpdfstring{$\alpha$}{alpha}-scans}

This note describes how the phase angles in \figref{fig:inversion_alpha} of the main text are obtained. The analysis
combines three steps, each corresponding to one element of the protocol in the main text: the
ellipticity from the eigenvalue envelope of an $\alpha$-averaged spectrum, the orientation of the
polarization ellipse from the $\alpha$-dependence of the central EIT peak, and the helicity from the
protocol of \fref{fig:handedness}. Angles are given in the convention of equation~(\ref{eq:phi}). The set phases are used only
to evaluate the result.

\subsection*{Data and analysis parameters.}
For the $50P_{3/2}\leftrightarrow49D_{5/2}$ transition ($J^p=\tfrac32^+$), $\alpha$-scans were
recorded at 100 RF phase angles spaced by $3.6^\circ$ over a full period. Each $\alpha$-scan consists
of 40 EIT spectra at antenna-stage angles from $0^\circ$ to $180^\circ$ in steps of $4.6^\circ$;
the spectrum at $180^\circ$ repeats the configuration at $0^\circ$ and is omitted, leaving 39
spectra per phase angle.

The line-detection parameters are those of Supplementary Note~\ref{sec:sup:inv} for $J^p=\tfrac32^+$.

\subsection*{Step 1: ellipticity.}
The 39 spectra of each $\alpha$-scan are averaged. Because the positions of the dressed-state
eigenvalues do not depend on $\alpha$ while the strengths of the individual lines do, the average
contains all lines of the eigenvalue pattern with reduced variations in their strengths. Lines are
detected and combined into pairs mirror symmetric about zero detuning. Two additional rules handle lines whose partner is not a separate maximum,
which occurs close to linear polarization where neighbouring eigenvalues are nearly degenerate.
(i) If a detected line at $\Lambda$ has no mirror partner, the partner is located by a least-squares
fit to the opposite side of the spectrum near $2c-\Lambda$ of two Lorentzians with a common width plus
a linear background, one describing the nearest detected line and one the expected partner (centre
free within the mirror-pair tolerance). The partner is accepted if its fitted amplitude exceeds three
times the residual r.m.s.\ and its separation from the neighbouring line exceeds the fitted width.
(ii) A detected line lying outside the outermost pair defines the outer envelope, its partner being
placed at the mirror position $2c-\Lambda$. Rule (i) was applied to 25 and rule (ii) to 1 of the 100
phase angles. The envelope ratio $R$, equation~(\ref{eq:ratio}), is formed from the innermost and outermost pairs
and inverted with equation~(\ref{eq:phitilde}), which gives $\phi_0=\arcsin|\sin\phi|\in[0,\pi/2]$. The exact
inversion from the eigenvalues of equation~(\ref{eq:big}) differs from equation~(\ref{eq:phitilde}) by at most $1^\circ$ for
these data.

\subsection*{Step 2: orientation of the polarization ellipse.}
For each of the 39 spectra of an $\alpha$-scan, the height $g$ of the central peak above the
spectrum's baseline is determined within the central zone. The peak will be minimized when the linear polarizations of the optical fields are along the major axis of the RF field's polarization ellipse (an extension of the results in Refs.~\onlinecite{Sedlacek2013} and \onlinecite{Cloutman2026} on co-polarized fields) and maximized when they are perpendicular to the major axis. Consequently, the peak will be maximal when the optical axis of the probe and coupling beam is along the ellipse's major axis. A linear least-squares fit to
\begin{equation*}
  g(\alpha)=A+B\cos\left[2(\alpha-\alpha_{\rm major})\right]
\end{equation*}
is used to determine $\alpha_{\rm major}$. Since the polarization ellipse is mirror symmetric about its axes,
$\alpha_\mathrm{major}$ coincides with an axis of the ellipse irrespective of the detailed shape of
$g(\alpha)$, and no calibration of the peak heights is required. On the prime meridian the major
axis is vertical for $|\phi|>\pi/2$ (the shaded region of Supplementary \fref{fig:envrat}) and horizontal for $|\phi|<\pi/2$. A spectrum is assigned to the
vertical case if $\cos2(\alpha_\mathrm{major}-135^\circ)>0$, where $135^\circ$ is the stage angle at
which the optical axis is horizontal in the rotated frame. This gives $|\phi|=\pi-\phi_0$ (vertical)
or $|\phi|=\phi_0$ (horizontal), i.e.\ the two candidates $\pm|\phi|$ shown in \figref{fig:inversion_alpha}a.

\subsection*{Step 3: helicity.}
The remaining ambiguity between $\phi$ and $-\phi$ is resolved with the helicity protocol of \fref{fig:handedness},
using the difference $\Delta^\circlearrowright_\circlearrowleft$ of the central-peak heights
measured in the configurations of \fref{fig:handedness}b,c, recorded on a separate grid of phase angles spaced by
$6^\circ$. Each phase angle of the $\alpha$-scans is assigned the sign of $\Delta$ at the nearest
measured phase angle, and $\mathrm{sgn}\,\phi=\mathrm{sgn}\,\Delta$.
Since $\Delta$ is odd in the ellipticity angle, it vanishes for linear polarization; its measured zero
crossing lies at $\phi\approx-11^\circ$.

\subsection*{Results.}
Supplementary Table~\ref{tab:fig10} summarises the deviations of the inferred from the set phase angle
for the 100 phase angles. For \figref{fig:inversion_alpha}a the deviation is evaluated for the nearer of the two candidates.
The orientation of the ellipse obtained in step~2 is correct for all 100 phase angles. The helicity
obtained in step~3 is ``wrong'' for 4 phase angles, all within $7^\circ$ of linear polarization
($\phi=-176^\circ,-173^\circ,-5^\circ,-2^\circ$), where $|\Delta|$ is small and the helicity is
ill-defined.

\begin{table}[h]
\centering
\caption{Deviation of the inferred from the set phase angle for the 100 $\alpha$-scans of \figref{fig:inversion_alpha}:
median, mean, r.m.s.\ and maximum deviation, and number of deviations exceeding $5^\circ$ and
$15^\circ$.}
\label{tab:fig10}
\begin{tabular}{lcccccc}
\toprule
 & Median & Mean & r.m.s. & Max. & $>5^\circ$ & $>15^\circ$\\
\midrule
\figref{fig:inversion_alpha}b (ellipticity and orientation)          & $2.0^\circ$ & $2.6^\circ$ & $3.2^\circ$ & $9.6^\circ$  & 15 & 0\\
\figref{fig:inversion_alpha}c (including helicity)                   & $2.2^\circ$ & $2.9^\circ$ & $4.0^\circ$ & $18.1^\circ$ & 19 & 1\\
\bottomrule
\end{tabular}
\end{table}

\subsection*{Remarks.}
The deviations in \figref{fig:inversion_alpha}b are not random but follow a systematic pattern: within
$|\phi|<\pi/2$ they mostly remain within $\pm3^\circ$ (maximum $4.5^\circ$), whereas for $|\phi|>\pi/2$ the magnitude
$|\phi|$ is underestimated by up to $\approx7^\circ$ (up to $10^\circ$ near circular polarization).
The pattern is identical for the closed-form and the exact inversion and therefore originates in the
measured line positions; it is consistent with the RF field deviating slightly from the prime
meridian, e.g.\ owing to a small amplitude imbalance between the two antenna ports or a non-linearity
of the phase setting. Compared with the single-configuration spectra of \figref{fig:inversion}b, averaging over
$\alpha$ restores lines of the innermost pair that are too weak to be detected in individual
spectra for one helicity, and the orientation of the ellipse is obtained from a symmetry argument
rather than from a threshold on the central-peak height.
\clearpage\newpage
\section{Inversion of independently acquired cesium spectra}\label{sec:sup:cs}

To showcase the universality of the inversion procedure, we applied it to Autler--Townes spectra recorded by Schlossberger \etal \cite{Schlossberger2026} on the $65D_{5/2}\leftrightarrow 66P_{3/2}$ transition in cesium, a realization of the $J^p=\tfrac{3}{2}^{+}$ member of the transition hierarchy in a different alkali species and at a different RF frequency ($\approx\SI{2.5}{GHz}$). We used the openly available data set (Ref.~21 of \citenum{Schlossberger2026}) recorded with parallel probe and coupling polarizations, comprising 400 EIT spectra for set ellipticity angles $\chi\in[-\pi/4,\pi/4]$. In the convention of Ref.~\onlinecite{Schlossberger2026} the RF field is $\mathbf{E}\propto\cos\chi\,\hat{\mathbf{z}}+i\sin\chi\,\hat{\mathbf{y}}$, so that $\chi$ is the Poincar\'e-sphere ellipticity angle, related to the phase angle of the present work by $|\sin\phi|=|\sin 2\chi|$.

\subsection*{Procedure}
Each spectrum was analysed independently, with the set values of $\chi$ used only to score the result. The analysis procedure follows Supplementary Note~\ref{sec:sup:inv}, with the following data-specific modifications. The detuning axis of the data file is rounded to three significant figures and was restored by a linear fit to the sample index (residuals $\le\SI{0.5}{MHz}$, consistent with the rounding). Both the line centre and the linewidth were obtained from a Lorentzian fit to the central line, instead of the Bare EIT spectrum, which was not available.

\subsection*{Results.}
\begin{figure*}[b!]
\includegraphics[width=0.9\textwidth]{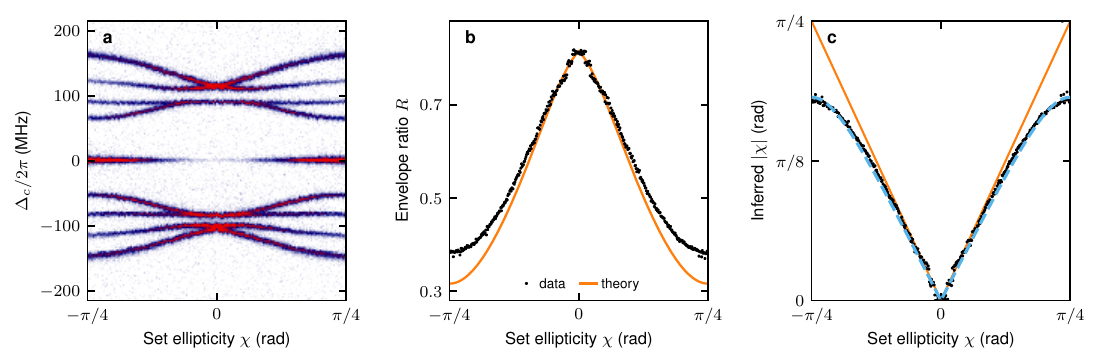}
\caption{\textbf{Inversion of independently acquired Cs spectra.} Data of Ref.~\citenum{Schlossberger2026} for the $65D_{5/2}\leftrightarrow 66P_{3/2}$ transition, recorded with parallel optical polarizations. \textbf{a}, EIT spectrogram as a function of the set ellipticity angle $\chi$, with the coupling detuning $\Delta_c$ measured from the line centre; red dots mark the innermost and outermost lines on each side identified by the per-spectrum peak search. \textbf{b}, Envelope ratio $R$ [Eq.~(\ref{eq:ratio})] (dots) compared with the value obtained from the eigenvalues of the coupling matrix (line). \textbf{c}, Ellipticity angle $|\chi|$ inferred with Eq.~(\ref{eq:phitilde}) (dots) versus set $\chi$. The expected inference for a non-quadrature phase difference is shown in dashed blue.}
\label{fig:S_NIST}
\end{figure*}

Supplementary \fref{fig:S_NIST} shows the identified envelope ratio, and the inferred ellipticity angle. At the linearly polarized setting, $|\chi|\le2^\circ$, the measured envelope ratio is $R=0.813\pm0.003$ (mean and standard deviation of 18 spectra), within 0.5\% of the value $\sqrt{2/3}\approx0.816$ fixed by angular-momentum algebra. For $|\chi|\le25^\circ$ (222 spectra) the inversion reproduces the set ellipticity angle with a median deviation of $0.9^\circ$ (r.m.s.\ $1^\circ$), without any adjustable parameter or calibration.

For $|\chi|>25^\circ$ the measured envelope ratio levels off at $R\approx0.38$, well above the value $R=\sqrt{1/10}\approx0.32$ expected for circular polarization, and the inferred ellipticity angle saturates at $|\chi|\approx32^\circ$ for both $\chi=-45^\circ$ and $\chi=+45^\circ$. This is not an artefact of the envelope-ratio inversion: a fit of all eight non-central line positions at the nominally circular settings to the eigenvalues of the coupling matrix, with the RF Rabi frequency, the line centre and $|\chi|$ as free parameters, yields $|\chi|=32.1^\circ$ and $32.2^\circ$ for $\chi=-45^\circ$ and $+45^\circ$, respectively, with r.m.s.\ residuals of \SI{2}{MHz}, whereas imposing $|\chi|=45^\circ$ doubles the r.m.s.\ residual and leaves deviations of up to \SI{11}{MHz}. The spectra recorded at the nominally circular settings are therefore self-consistently those of an elliptically polarized field with $|\chi|\approx32^\circ$, indicating that the RF field at the atoms did not reach circular polarization. The dashed blue line on Supplementary \fref{fig:S_NIST}c shows the expected behaviour if the signals used to generate the elliptical RF field are separated by $66.5^\circ$ instead of $90^\circ$.

\newpage
\renewcommand{\refname}{\textbf{Supplementary References}}
\putbib
\end{bibunit}

\makeatletter
\let\bibliography\@gobble
\makeatother
\end{document}